\documentclass[lettersize,journal, twoside]{IEEEtran}
\usepackage{amsmath,amssymb,amsfonts}
\usepackage{cite}
\usepackage{epsfig}
\usepackage{graphicx}
\usepackage{epstopdf}
\usepackage{algorithmic}
\usepackage{textcomp}
\usepackage{xcolor}
\usepackage{color}
\usepackage{soul}
\usepackage[caption=false,font=footnotesize]{subfig}
\usepackage{xfrac}
\usepackage[hyphens,spaces,obeyspaces]{url}
\usepackage{bbm}
\usepackage{dsfont}
\usepackage{comment}
\usepackage{booktabs}
\usepackage{multirow}
\usepackage{array} 
\usepackage[colorlinks=true,
            linkcolor=blue,
            urlcolor=black,
            citecolor=blue,
            anchorcolor=blue]{hyperref}

\ifodd 0
  \newcommand{\CommentWong}[1]{\textcolor[rgb]{1,0,0}{[Wong comment: #1]}}
  \newcommand{\EditPrasun}[1]{\textcolor[rgb]{0,0,1}{#1}}
  \newcommand{\CommentPrasun}[1]{\textcolor[rgb]{0,0,1}{[Prasun comment: #1]}}

  \newcommand{\hlb}[1]{\textcolor{blue}{#1}}
  
\else
  \newcommand{\CommentWong}[1]{}
  \newcommand{\CommentPrasun}[1]{}
  \newcommand{\commentAni}[1]{}  
  \newcommand{\EditPrasun}[1]{#1}
  
  \newcommand{\hlb}[1]{}
  
\fi

\graphicspath{{./figs/}}

\def \p{\boldsymbol{\mathrm{p}}}
\def \n{\boldsymbol{\mathrm{n}}}
\def \v{\boldsymbol{\mathrm{v}}}
\def \X{\boldsymbol{\mathrm{X}}}

\def \o{\boldsymbol{\mathrm{o}}}

\def \H{{\bf H}}

\def \S{{\bf N}_{\text{s}}}
\def \C{{\bf N}_{\text{c}}}

\def\Srm{S^\text{rm}}
\def\Tmax{T^\text{max}}
\def\Tc{T^\text{r}}

\begin{document}


\title{Chip-Surface Based Visual Authentication \\for Integrated Circuits}




\author{Runze Liu, 
Prasun Datta,$^*$~\IEEEmembership{Graduate Student Member,~IEEE,} 
Anirudh Nakra,$^*$~\IEEEmembership{Graduate Student Member,~IEEE,}
Chau-Wai~Wong,~\IEEEmembership{Senior Member,~IEEE,}
and~Min~Wu,~\IEEEmembership{Fellow,~IEEE}%
\thanks{* Equal contributions. This work was supported in part by the US National Science Foundation (award numbers ECCS-2227499 and ECCS-2227261 \textit{(Corresponding author: Chau-Wai Wong}).}%
\thanks{Runze Liu is now an independent researcher. He conducted this research work while he was with the Department of Electrical and Computer Engineering, NC State University, NC 27695 USA.}%
\thanks{Prasun Datta and Chau-Wai Wong are with the Department of Electrical and Computer Engineering, the Forensic Science Cluster, and the Secure Computing Institute, NC State University, NC 27695 USA.}%
\thanks{Anirudh Nakra and Min Wu are with the Department of Electrical and Computer
Engineering and the Institute for Advanced Computer Studies, University of Maryland, College Park, MD 20742 USA.}
}

\maketitle

\begin{abstract}
The rapid development of the semiconductor industry and the ubiquity of electronic devices have led to a significant increase in the counterfeiting of integrated circuits~(ICs). 
This poses a major threat to public health, the banking industry, and military defense sectors that are heavily reliant on electronic systems.
The electronic physically unclonable functions~(PUFs) are widely used to authenticate IC chips at the unit level. 
However, electronic PUFs are limited by their requirement for IC chips to be in working status for measurements and their sensitivity to environmental variations. 
This paper proposes using optical PUFs for IC chip authentication by leveraging the unique microscopic structures of the packaging surface of individual IC chips. 
The proposed method relies on color images of IC chip surfaces acquired using a flatbed scanner or mobile camera. 
Our initial study reveals that these consumer-grade imaging devices can capture meaningful physical features from IC chip surfaces. 
We then propose an efficient, lightweight verification scheme leveraging specular-reflection-based features extracted from videos, achieving an equal error rate~(EER) of 0.0008. 
We conducted factor, sensitivity, and ablation studies to understand the detailed characteristics of the proposed lightweight verification scheme.
Our work is the first to apply the optical PUF principle for the authentication of IC chips, synergizing image and video processing with semiconductor chip technology and demonstrating the potential to significantly enhance the security of the semiconductor supply chain.
\end{abstract}


\section{Introduction}
The semiconductor industry has been developing rapidly over the past few decades, and electronic devices are now an integral part of our daily lives.
In recent years, the counterfeiting of integrated circuit~(IC) chips has become a significant challenge due to the restructuring of global supply chains. 
The use of counterfeit IC chips poses threats to various sectors that heavily rely on electronic systems, such as public health, the banking industry, and military defense. 
According to the Semiconductor Industry Association, counterfeiting costs US-based semiconductor companies more than \$7.5 billion annually and results in the loss of nearly 11,000 jobs~\cite{villasenor2013chop}. 
Unreliable counterfeit IC chips have even led to fatal real-life incidents~\cite{IC_Safety_Death}. 
The intermittent shortage of IC chip supplies has forced supply-chain participants to procure IC chips from unreliable sources, thereby increasing the risk of acquiring counterfeit IC chips.
Consequently, developing effective and efficient anti-counterfeiting techniques for IC chips has become increasingly important.

Electronic physically unclonable functions~(PUFs) have been used for anti-counterfeiting for ICs~\cite{roel2012physically}. 
Due to the manufacturing variations of ICs, the electronic measurements of each device, such as voltages, resistance, digital time delays, and power-up states of cells are unique and unpredictable. 
Such manufacturing variations are impossible to duplicate. 
However, electronic PUFs require the ICs to be put into working status to obtain the measurements, which usually needs trained personnel to operate them in a working laboratory environment.
In a real-world scenario, such as in a supply chain, it is desirable to obtain measurements conveniently.
The electronic PUFs have also been shown to be sensitive to aging and environmental variations, such as thermal noise and power supply noise \cite{gao2020physical}. Such disadvantages of electronic PUFs make them less effective in real-world applications for anti-counterfeiting for IC chips.

The surfaces of objects are random and uneven due to their unique microscopic structures (microstructures), which can be regarded as their ``fingerprints.''
Optical PUF was first proposed to identify the unique, three-dimensional~(3-D) microstructure of a plastic object using laser speckles~\cite{pappu2002physical}.
Later, optical PUFs were successfully used for the identification of paper surfaces in various applications, including product authentication, document forgery prevention, and counterfeit drug detection~\cite{buchanan2005forgery, clarkson-09, voloshynovskiy2012towards, diephuis2013physical, diephuis2014framework, wong2017counterfeit, liu2021microstructure,datta2024enabling, chaban2024comparative, beekhof2008secure,sharma2011paperspeckle,toreini2017texture,kauba2016towards,schraml2018real}. Since the surfaces of the packaging of IC chips are also random like paper surfaces due to manufacturing variations, we hypothesize that optical PUFs can also be used for the unique authentication of IC chips.
Unlike electronic PUFs that are sensitive to temperature changes and power supply noise, the optical ``responses'' of an IC chip surface to the visual light are more stable.
Furthermore, the acquisition of an optical response is fast when using optical imaging devices such as a flatbed scanner or camera, which is an advantage for verification in supply-chain applications. 
Also, advances in computer vision and photo acquisition devices lower the potential barriers to exploiting optical PUFs for IC chip authentication. 
This paper will also address such technical challenges as registering captured images of chip surfaces and the design of an efficient, lightweight verification scheme.

In this work, we propose using the optical PUF principle to authenticate IC chips, where we capture photos of the packaging of an IC chip with a flatbed scanner or mobile camera to obtain the physical features of the chip surface.
To the best of our knowledge, this is the first work to apply the optical PUF principle for IC chip authentication.
This IC chip authentication paper comprises three main efforts. 
First, we investigate whether the images captured with consumer-grade imaging devices can provide sufficient details about the physical characteristics of IC surfaces with reference to measurements from a confocal microscope. 
Second, we propose a video-based, fast verification method that leverages specular-reflection-based features for authentication.
Third, we perform ablation studies to assess the impact of edge and masking techniques and conduct factor analyses of frame selection and specular points.
The contributions of this work are threefold.
\begin{itemize}
  \item Our physics-related explorations confirm that consumer-grade imaging devices such as cameras can capture sufficient physical surface details for IC chip authentication.
  \item We show that specular points can better serve as authentication features than norm maps or height maps.
  Utilizing specular points offers reduced error rates, computational complexity, and communication overhead, making it more feasible for real-world deployment.
  \item Our extensive experimental evaluations provide a guide to the optimal mode of operation for the proposed specular-point-based authentication method.
\end{itemize} 

Our work leverages image/video processing, security, and semiconductor circuits and systems to introduce counterfeit-prevention mechanisms for chips---an increasingly critical direction in the era of computation.

\section{Related Work}
\label{sec:related-work}
\subsection{Electronic PUFs for ICs}
Electronic PUFs can be used for IC identification by utilizing the inherent manufacturing variations in ICs.
These variations manifest in diverse physical properties such as voltages, resistances, and digital delays. These properties provide unique, unclonable ``fingerprints'' for the authentication of chips. For instance, Lofstrom et al.~\cite{lofstrom2000ic} exploited the threshold voltage variations of transistors to create a PUF system for IC authentication. Helinski et al.~\cite{helinski2009physical} harnessed the resistance variations in power grids caused by manufacturing inconsistencies as an authentication feature. Besides analog measurements, digital delay-based PUFs were proposed to improve IC authentication. Gassend et al.~\cite{gassend2002silicon} introduced the arbiter PUF, which utilizes statistical variations in device and wire delays within the IC. Lee et al.~\cite{lee2004technique} and Patel et al.~\cite{patel2009increasing} extended the delay-based approach to incorporate transistors and combinational circuits to generate secret keys and authenticate IC chips.

Another line of work focuses on destabilized memory cells as a PUF structure. Holcomb et al.~\cite{holcomb2007initial} demonstrated the use of static random-access memory (SRAM) cells as a fingerprinting mechanism by utilizing their unique power-up states. This concept was later expanded to include other storage elements, such as flip-flops~\cite{maes2008intrinsic} and buskeeper cells~\cite{simons2012buskeeper}, which diversify the range of memory-based PUF solutions. However, despite their advantages, these electronic PUFs are highly sensitive to environmental factors such as temperature fluctuations and power supply noise. This makes their deployment in practical scenarios more challenging. While each of these methods provides a valuable framework for IC authentication, their practicality is limited by the reliance on electronic measurements and the requirement for the IC to be in working status. The sensitivity of electronic PUFs to environmental variations, such as thermal noise, introduces additional challenges. This highlights the potential for complementary approaches, e.g., optical PUFs that leverage the physical surface properties of IC chips, as will be discussed in the next subsection.

\subsection{Optical PUF for Paper Surfaces}
Optical PUFs for paper surface identification may be categorized generally into two types: optical feature-based methods and physical feature-based methods. The optical feature-based methods focus on leveraging the visual appearance of the paper surface under the light for authentication. For example, Buchanan et al.~\cite{buchanan2005forgery} demonstrated the potential of optical PUFs by scanning paper surfaces with a laser beam and analyzing the resulting laser speckle patterns for authentication purposes. Beekhof et al.~\cite{beekhof2008secure} proposed a more accessible method by using macro-lens-equipped mobile phones to capture paper surface images. They proposed using minimum reference distance and reference list decoding for accurate authentication. Sharma et al.~\cite{sharma2011paperspeckle} utilized a camera coupled with a microscope and built-in LED to capture fine paper surface speckles as unique fingerprints for paper authentication. 
Toreini et al.~\cite{toreini2017texture}, instead of capturing reflected light from a paper surface, propose capturing the visual image of light that transmits through the paper texture.

The proposed authentication of IC chips using surfaces was inspired by prior studies that utilized physical features of unique microstructures of paper surfaces for authentication purposes. The norm map, a projection of uniformly spaced surface normals onto the $x$-$y$ plane, can be used to quantify the microstructures of a surface. Clarkson et al.~\cite{clarkson-09} proposed a method for estimating a scaled version of the norm map by scanning paper patches in opposite directions with a flatbed scanner, assuming light reflection is fully diffuse. This method laid the foundation for the use of physical features for authentication. Wong and Wu~\cite{wong2015study,wong2015counterfeit,wong2017counterfeit} proposed an authentication method that uses a mobile camera with a built-in flash to capture multiple images of a paper patch from different perspectives. By applying a diffuse reflection model in conjunction with the camera's geometry, they estimated the norm map, which was further validated against ground-truth data obtained through a confocal microscope. Their method provided a portable and accessible solution, enhancing the practicality of paper surface-based authentication. Liu et al.~\cite{liu2018enhanced} introduced improvements to the norm map estimation process by accounting for factors such as ambient light and the camera's internal brightness and contrast adjustments. Their work highlighted the importance of high-spatial-frequency components of height maps, offering better authentication performance than norm maps. 
Section~\ref{sec:section-diffuse} will apply diffuse-based norm map estimation to IC chip surfaces and confirm its feasibility with confocal scans.
Liu and Wong~\cite{liu2021microstructure} also showed that specular reflection is not an important factor for paper surface authentication in the flatbed scanner setup. While optical PUFs offer state-of-the-art authentication performance for paper surfaces, adapting these methods for IC chip surfaces presents new challenges, especially in handling specular reflection. This requires a customized authentication method that exploits the specular properties of IC chip surfaces, which will be detailed in Section~\ref{sec:fast-acc-auth-spec}.

\subsection{Encapsulation for IC Chips} 
\label{sec:IC_encapsulation}

\begin{figure}[!t]
\centering
  \vspace{-1mm}
  \hspace{-1mm}
  \subfloat[]{\includegraphics[width=0.48\linewidth]{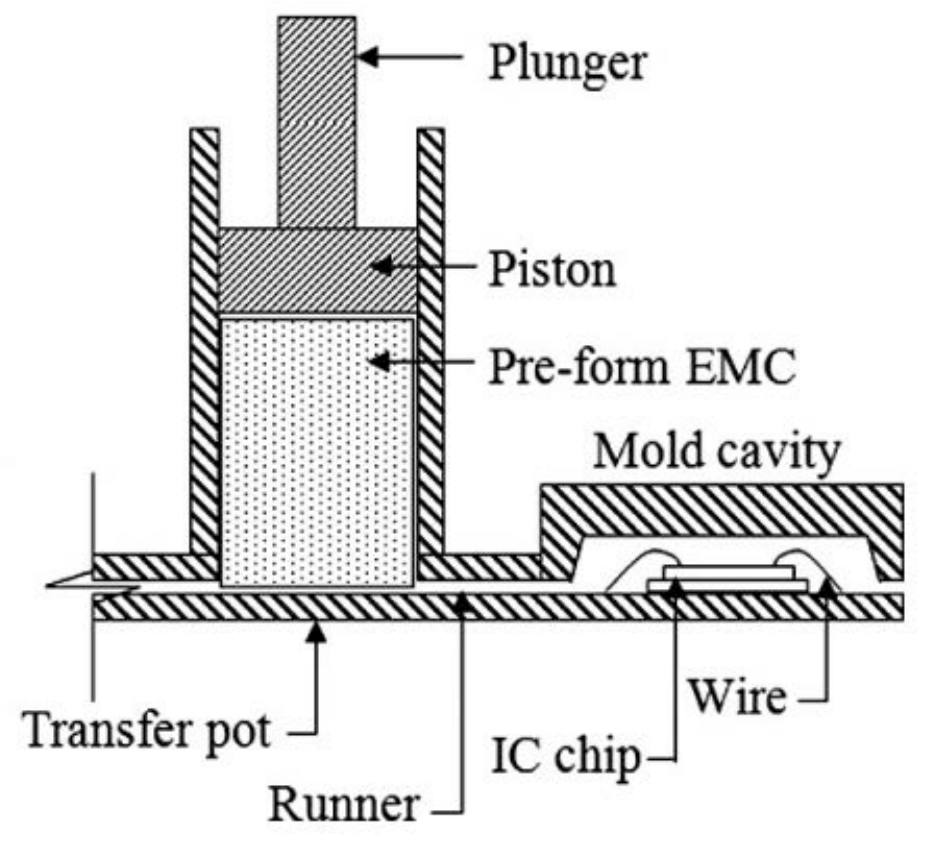}}
  \subfloat[]{\includegraphics[width=0.48\linewidth]{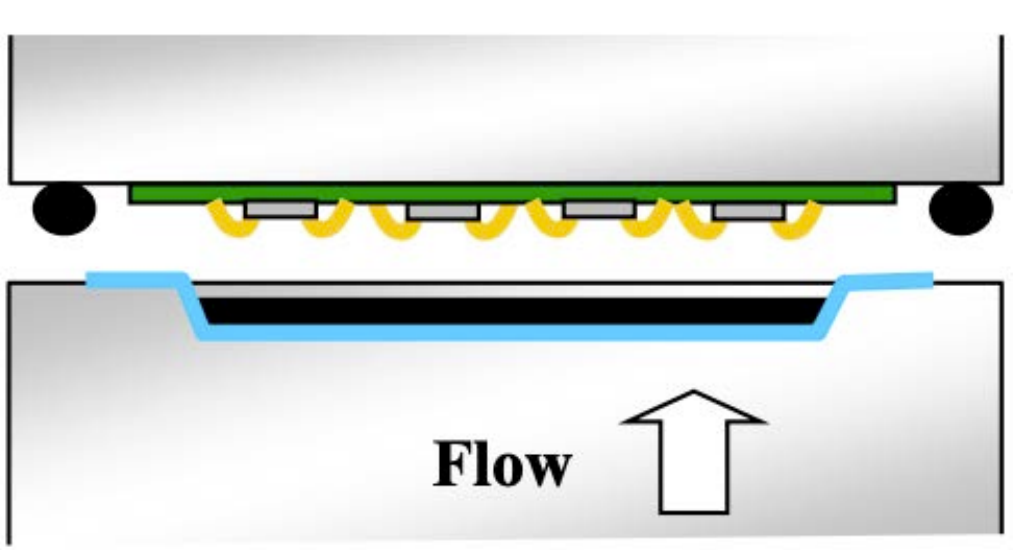}}
  \hspace{2mm}
  \caption{(a)~An illustration of the transfer molding encapsulation setup~(reproduced from~\cite{khor2014recent}). The pre-formed EMC will be transferred from the transfer pot via the runner to the mold cavity. The IC will be encapsulated and protected by the EMC. (b)~An illustration of compression molding~(reproduced from~\cite{miura2016compression}). The mold with compound will be closed by applying the required pressure, with a vacuum to suck up air, gas, and moisture coming out from the compound. These encapsulation processes lead to unique identifiers on the IC packaging surfaces at the microscopic level, which may be used for IC chip authentication.
  }
  \label{fig:encapsulation}
\end{figure}

Integrated circuits~(ICs) are typically encapsulated by epoxy-molding compound (EMC) material to protect against moisture, ionic contamination, and thermal and mechanical threats~\cite{khor2014recent}. The resulting IC packaging surfaces may provide sufficient microscopic details for authentication purposes. We review two main IC encapsulation methods. \textit{Transfer molding}~\cite{khor2014recent}, as illustrated in Fig.~\ref{fig:encapsulation}(a), involves using liquefied epoxy resin to submerge the IC in a mold cavity. The surfaces of IC chip packaging are formed during the encapsulation process and exhibit random unevenness due to flow marks, rough mold cavities, and air traps~\cite{siong2018effect}. 
The hardening of the resin during the curing process may further contribute to surface randomness, leading to unique microstructural identifiers that can be used for IC chip authentication purposes. \textit{Compression molding}~\cite{miura2016compression}, as illustrated in Fig.~\ref{fig:encapsulation}(b), uses a granular compound rather than liquefied resin by applying pressure to create IC packaging. A random packaging surface forms during the curing stage. Unlike paper surfaces, these IC packaging surfaces contain more specular points due to the nature of the epoxy material. 
These specular points are utilized in the fast verification method proposed in Section~\ref{sec:fast-acc-auth-spec}.

\section{Investigating Diffuse Reflection Based Features for IC Chips}
\label{sec:section-diffuse}

The randomness of IC chip surfaces can potentially serve as intrinsic fingerprints for the unique identification of IC chips.  
In this section, we examine whether consumer-grade imaging devices can capture sufficient microscopic details of the surfaces of IC chips.
We first examine the feasibility of extracting meaningful microstructures using flatbed scanners, then extend to using mobile cameras, and finally assess the authentication performance of the extracted microstructures.

\subsection{Feasibility of Capturing Microstructures of IC Chip Surfaces Using a Consumer-Grade Flatbed Scanner}
\label{sec:scanner-diffuse}

We first explore the possibility of 
extracting
meaningful physical features 
of IC chip surfaces
using flatbed scanners. 
The use of
scanners instead of cameras 
allows us to acquire higher-quality imaging signals.
We compare the norm maps estimated from scanner-captured images with those obtained from a confocal microscope, which 
we consider
as the ground truth 
given the confocal microscope's capabilities to accurately measure the height maps of the chip surfaces.

We used a CanoScan LiDE 300 flatbed scanner as in \cite{liu2021microstructure} to scan the images of a chip surface and obtain its norm map.
Following the norm map estimation algorithm for surfaces with diffuse and specular components in Appendix~\ref{sec:norm-map-est-alg}, we scanned each IC chip surface in two pairs of opposite directions and took the difference between the scanned images of each pair as in \eqref{eq:scanner_diff} to 
extract the scaled
$x$- or $y$-component of the norm map.
The scanner's resolution was $600$ pixels per inch~(ppi), translating to a pixel edge length of $42.33$ $\mu$m.
To compute the ground-truth norm map, we used a Keyence VKx1100 confocal microscope to measure the height map of the same chip surface.
The pixel edge length was set to be $5.37$ $\mu$m.

We assess the 
quality of the norm maps extracted from the scanner with reference to those from the confocal microscope.
We selected a part of the scanner's norm map 
$\S$, 
with the size of $70\times70$ pixels 
from
the background region of the chip surface 
without texts.
We did so because identical foreground texts on different chip units may exhibit identical engraved strokes on the chip surfaces, leading to similar norm maps and potential false positives during authentication.
To facilitate a 2-D cross-correlation search of the norm map over the confocal norm map,
we upsampled the confocal height map to obtain 
a modified height map 
$\H$, such that the edge length of the confocal height map is $2.65$ $\mu$m, i.e., $1/16$ of that of the scanner's pixel edge length.
We then searched for an area $\H_0$, of the same physical size as $\S$, within $\H$ such that the norm map $\C$ obtained from $\H_0$ exhibits the highest cross-correlation value with $\S$. We followed the procedure in \cite{wong2017counterfeit} to derive the norm map $\C$ from the heightmap $\H_0$.

\begin{figure}[!t]
\centering
  \vspace{-1mm}
  \hspace{-4mm}
  \subfloat[]{\includegraphics[height=0.156\textheight,width=0.48\linewidth]{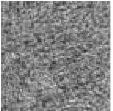}}
  \hfil
  \subfloat[]{\includegraphics[width=0.49\linewidth,trim={5mm 5mm 0 0},clip]{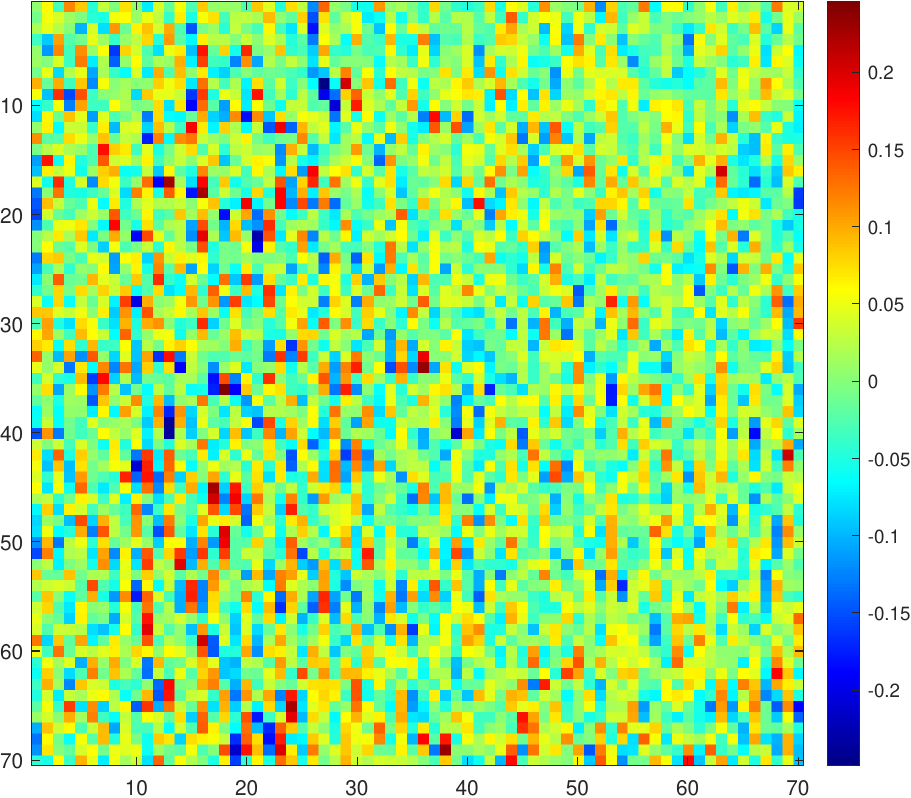}}
  \hspace{-4mm}

  \hspace{-4mm}
  \subfloat[]{\includegraphics[width=0.48\linewidth,trim={10mm 5mm 0 0},clip]{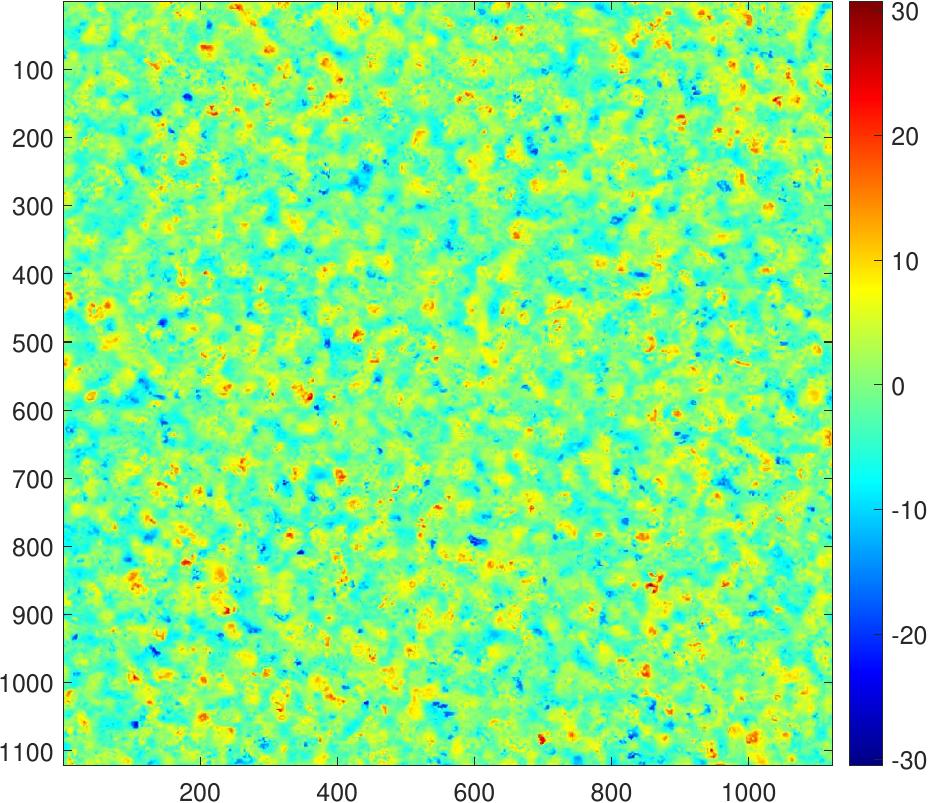}}
  \hfil
  \subfloat[]{\includegraphics[width=0.48\linewidth,trim={5mm 5mm 0 0},clip]{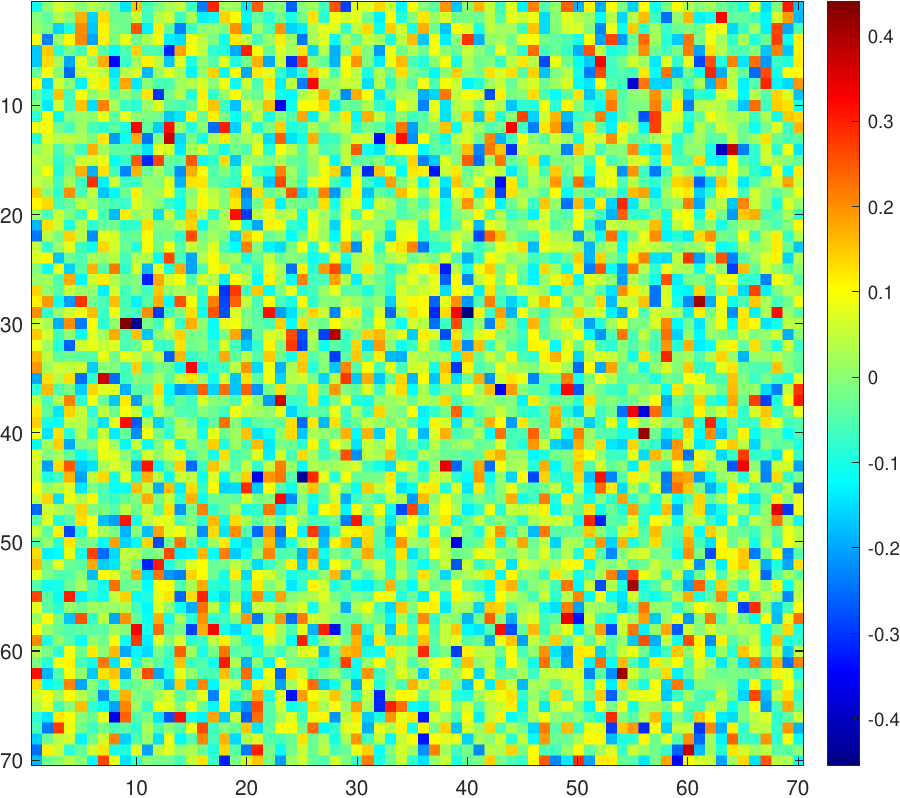}}
  \hspace{-3.6mm}
  \caption{(a) A scanner-captured image (after contrast enhancement) of an area in the background of IC chip surface. (b) The estimated $x$-component of norm map $\S$ from the scanner. (c) The height map $\H_0$ measured by confocal microscope and (d) the derived norm map $\C$ from $\H_0$.}
  \label{fig:scanner_confocal_consistency}
\end{figure}

We evaluated the cross-correlation between $\C$ and $\S$ on two different chips, selecting three different 70-by-70-pixel background regions on each chip.
When $\S$ and $\H$ are from the same chip surface, the correlation between $\C$ and $\S$ should be high. 
Indeed, in our experiments, the cross-correlations were 0.53, 0.53, and 0.54.
An example of the estimated norm map $\S$, confocal heightmap $\H_0$, and confocal norm map $\C$ for a chip is shown in Fig.~\ref{fig:scanner_confocal_consistency}.
In contrast, when $\S$ and $\H$ were from two different chip surfaces, the cross-correlations between $\C$ and $\S$ dropped to 0.04, 0.04, and 0.03.
These results demonstrate that consumer-grade imaging devices can effectively capture meaningful microstructures of the chip surface.

\begin{figure}[!t]
\centering
  \vspace{-1mm}
  \hspace{-1mm}
  {\includegraphics[width=0.9\linewidth]{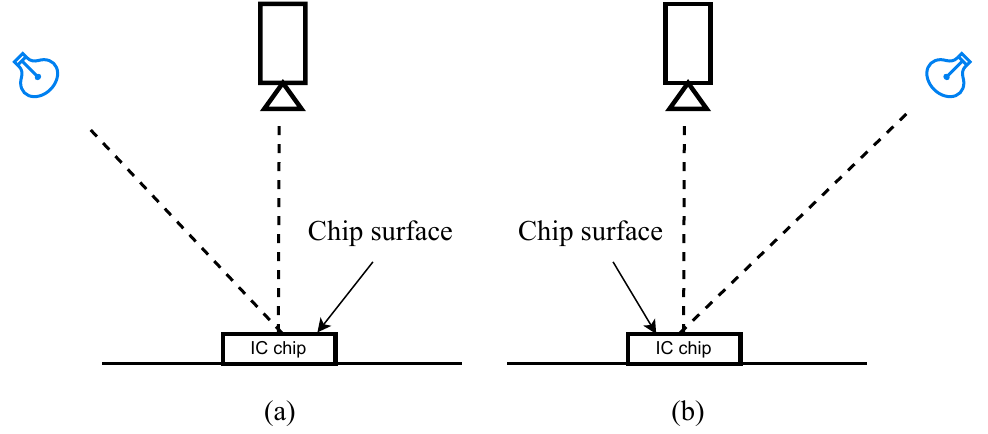}}
	\caption{Two side views when using a camera to capture images of IC chip surface with a light source, aiming to mimic the scanner setups of placing the scanned item at (a) ${0}^\circ$ and (b) ${180}^\circ$ so that scanner's norm map estimation algorithm can still be applied. Taking the difference between the two captured images results in a scaled version of the $x$-component of the norm map.}
  \label{fig:capture_IC_image}
\end{figure}

\subsection{Ubiquitous Capture of Microstructures of IC Chip Surfaces Using Mobile Cameras}
\label{subsec:auth-feature-diffuse}

Section~\ref{sec:scanner-diffuse} demonstrates that flatbed scanners can capture meaningful 3-D microstructures of IC chip surfaces for unique identification purposes. However, scanners are bulky and not widely available, whereas cameras are more ubiquitous. We now investigate the potential of using mobile cameras to capture the physical features of IC chip surfaces.

\vspace{1mm}\noindent{}\textbf{Imaging Setup.} Fig.~\ref{fig:capture_IC_image} illustrates the experimental setup that mirrors that of the scanner in Section~\ref{sec:scanner-diffuse} so that the same norm map estimation algorithm for scanners is valid to be used for mobile cameras. We positioned an iPad Pro 2018 camera directly above the IC chip surface. Denoting the center of the IC chip surface as the origin, a lamp was placed such that the polar angle of the incident light direction is approximately $45^\circ$ from the vertical axis. We captured three sets of images at each of the following azimuthal angles of the lamp: ${0}^\circ$, ${90}^\circ$, ${180}^\circ$, and ${270}^\circ$, resulting in a total of 12 images per chip.

\begin{figure}[!t]
\centering
  \vspace{-1mm}
  \hspace{-1mm}
  \subfloat[]{\includegraphics[width=0.48\linewidth]{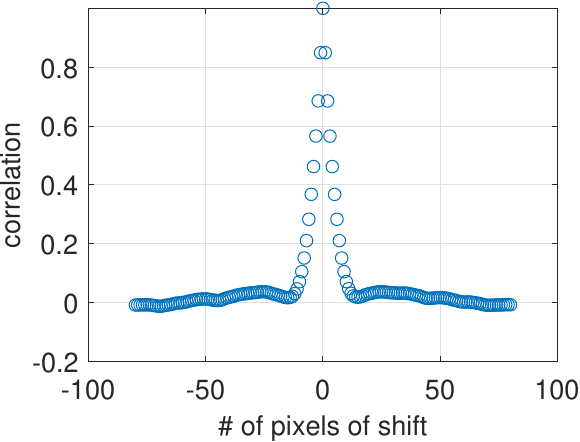}}
  \hspace{2mm}
  \subfloat[]{\includegraphics[width=0.48\linewidth]{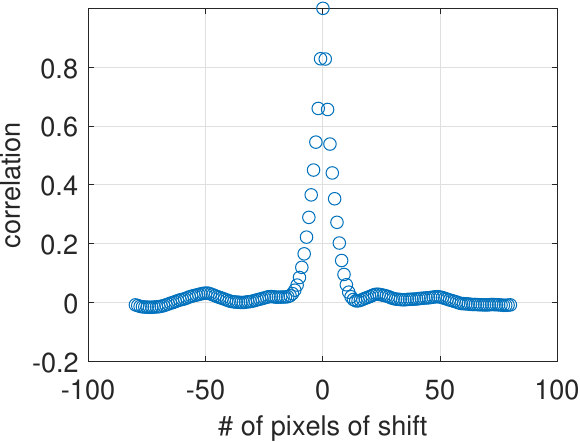}}
	\caption{The spatial autocorrelation values of a height map with shifts in the (a)~$x$- and (b)~$y$-directions. IC chip's correlations drop below 0.5 with merely 4 confocal-pixel shifts and below 0.2 with 8 shifts.}
  \label{fig:heightmap_shift}
\end{figure}

\begin{figure}[!t]
\centering
  \vspace{-1mm}
  \hspace{-2.6mm}
  \subfloat[]{\includegraphics[height=0.096\textheight,width=0.32\linewidth]{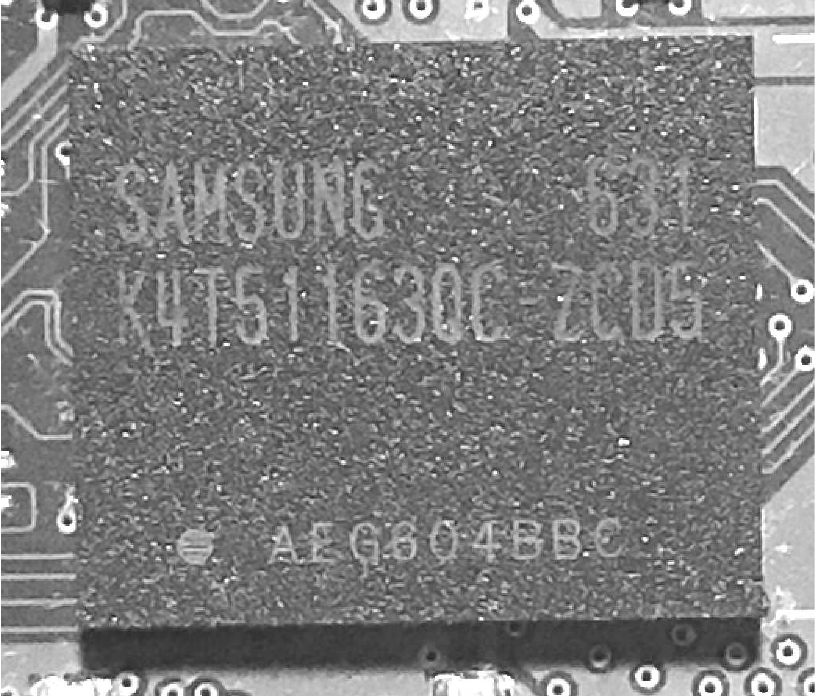}}
  \hfil
  \subfloat[]{\includegraphics[width=0.32\linewidth]{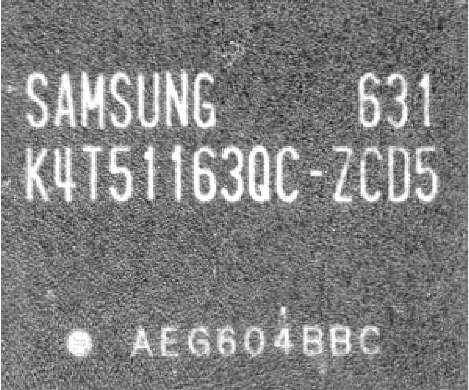}}
  \hfil
  \subfloat[]{\includegraphics[width=0.32\linewidth]{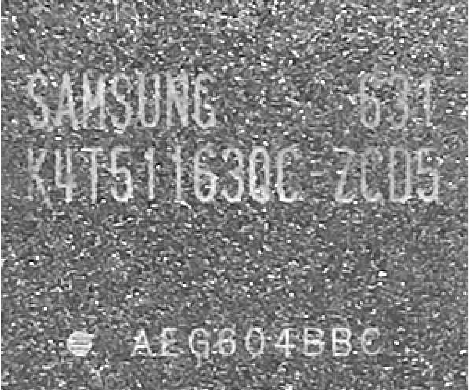}}
  \hspace{-1.5mm}
  \caption{An example of (a) a camera-captured image, (b) the template image used, and (c) the registered camera image. All images have undergone contrast enhancement for better visualization.}
  \label{fig:registered_image}
  \vspace{-4mm}
\end{figure}

\vspace{1mm}\noindent{}\textbf{Precise Image Alignment.}
We developed a precise alignment algorithm for IC chip images due to the small scale of the microstructures on IC chip packaging surfaces.
We assessed the scale of the microscopic features using the spatial autocorrelation of a height map as a proxy.
The height map of a background region measuring $8.78$~mm$^2$ was acquired using a confocal microscope.
The blue circles in Fig.~\ref{fig:heightmap_shift}(a) and (b) represent the correlation coefficients as functions of horizontal and vertical shifts, respectively. 
Both plots reveal that with a misalignment of 4 and 8 confocal pixels, the correlation drops to below 0.5 and 0.2, respectively.
Given that the pixel length in the scanner is approximately eight confocal pixels as stated in the previous subsection, this implies that a half- or full-pixel misalignment, without considering other error sources, could significantly lower the matching score for true positive cases.
We therefore proposed an alignment algorithm in Appendix~\ref{subsec:alignment-phase} to achieve the required accuracy for aligning IC chip images.
An example of a camera-captured image before and after alignment is shown in Fig.~\ref{fig:registered_image}.

\vspace{1mm}\noindent{}\textbf{Authentication Performance Using Optical PUFs.}
We examine the authentication performance when using such physical features as the norm map and the height map for IC chip authentication.
The test images are captured by a mobile camera, which is ubiquitous and user-friendly for real-world deployment.
The reference images are captured by a flatbed scanner for higher quality images with less noise compared to mobile cameras.

\begin{figure}[!t]
\centering
  \vspace{-1mm}
  \hspace{-1mm}
  \subfloat[]{\includegraphics[width=0.48\linewidth]{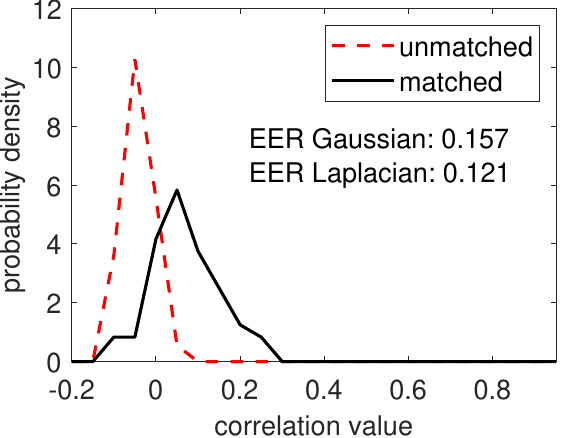}}
  \hspace{2mm}
  \subfloat[]{\includegraphics[width=0.48\linewidth]{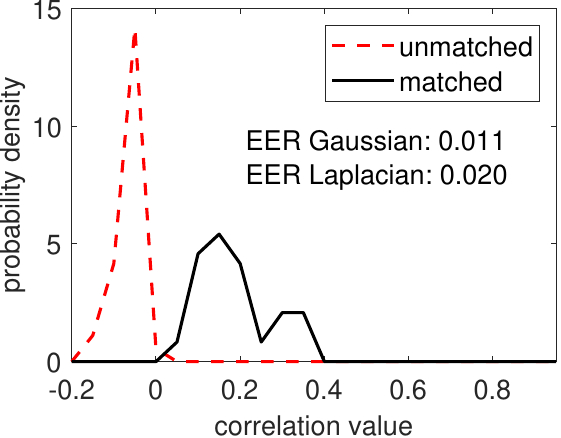}}
  \caption{Histograms of correlation values between (a) $x$- or (b) $y$-component of norm maps estimated from the mobile camera and flatbed scanner measurements. The discriminative performance is much better when using the $y$-component of norm maps.}  
  \label{fig:camera-scanner_nm}
\end{figure}
We first use the norm map as the authentication feature and use the correlation between the test and reference as the test statistic. Fig.~\ref{fig:camera-scanner_nm}(a) shows the estimated probability density functions~(PDFs) of correlation for the matched and unmatched cases, when using the $x$-component of the norm map as the authentication feature. The correlation values are higher under the matched cases, indicating that norm maps estimated from the camera may be used for IC authentication. Quantitatively, the EER is 0.12 when fitting Laplace PDFs~(and 0.16 when fitting Gaussian PDFs). When using the $y$-component as the feature, Fig.~\ref{fig:camera-scanner_nm}(b) shows a larger distance between the two histograms, indicating an improved discriminative capability with EER of 0.02 for Laplace PDFs~(and 0.01 for Gaussian PDFs). The better performance of the $y$-component may be rooted in the transfer molding process, in which the material flows in a specific direction that favors it. When the pre-EMC is transferred in the $x$-direction, the rough cavity may leave horizontal traces, which can be reflected in the $y$-component of the norm map. We include in the second row of TABLE~\ref{tab:comp-main} the best possible Laplacian EER achieved by the norm map as a baseline to highlight the effectiveness of the proposed method in Section~\ref{sec:fast-acc-auth-spec}.
\begin{table}[!t]
  \caption{Performance of Various Features for IC Authentication \vspace{-2mm}}
  \centering
  \scalebox{1.0}{
  \begin{tabular}{llll}
  \toprule
  \textbf{Sensing} & \textbf{Authentication} & \textbf{Test} &  \textbf{Laplacian}  \\ 
  \textbf{Modality} & \textbf{Feature} & \textbf{Statistic} & \textbf{EER} \\
  \hline
  Image & Intensity (Sec.~\ref{subsec:auth-feature-diffuse}) & Correlation &   $5 \times 10^{-2}$ \\
  Image & Norm map (Sec.~\ref{subsec:auth-feature-diffuse}) & Correlation &   $2 \times 10^{-2}$ \\
  Image & Height map (Sec.~\ref{subsec:auth-feature-diffuse}) & Correlation  & $5 \times 10^{-3}$ \\
  Video & Raw frames (Sec.~\ref{subsec:auth-feature-video-frames})  &  Max-correlation & $3 \times 10^{-4}$ \\
  Video & Specular points (Sec.~\ref{sec:fast-acc-auth-spec})  & Score $\Tc$ & $8 \times 10^{-4}$ \\
  \bottomrule
  \end{tabular}}
  \label{tab:comp-main}
  \vspace{-1mm}
\end{table}

\begin{figure}[!t]
\centering
  \vspace{-1mm}
  \hspace{-1mm}
  \subfloat[]{\includegraphics[width=0.48\linewidth]{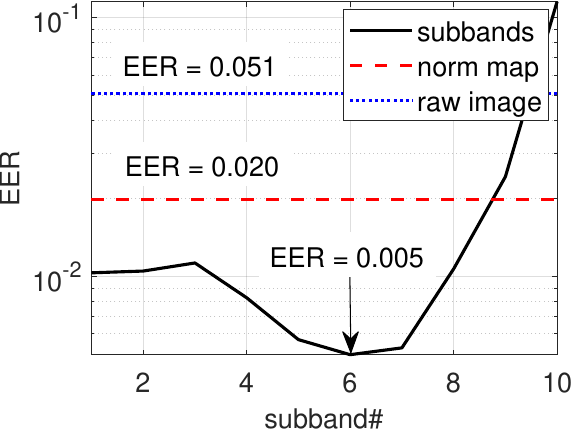}}
  \hspace{2mm}
  \subfloat[]{\includegraphics[width=0.48\linewidth]{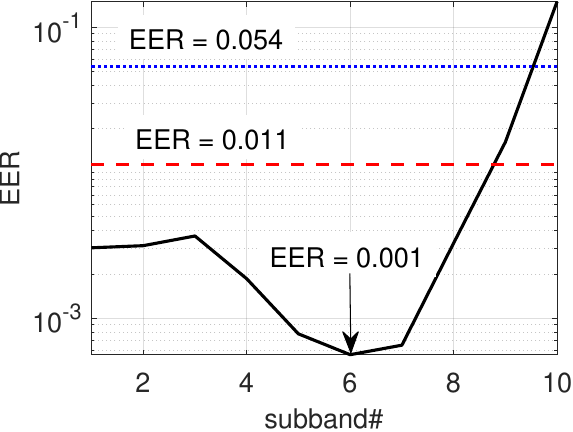}}
  \caption{EERs for different subbands of height maps used as the authentication feature when correlation values are believed to follow (a)~Laplace or (b)~Gaussian PDFs. 
  The sixth subband is the most discriminative physical feature, achieving the lowest EER of $5 \times 10^{-3}$ 
  under the Laplacian assumption and $6 \times 10^{-4}$ under the Gaussian assumption.
  The horizontal lines indicate the weaker baselines when using the $y$-component of the norm map or the raw images.
  }
  \label{fig:compare_physical_features_cam_sca}
\end{figure}

We also tested the state-of-the-art authentication feature, the spatial-frequency subband of height map~\cite{liu2018enhanced, liu2021microstructure}, as the IC chip authentication feature. Different spatial-frequency subbands of the reconstructed height map can capture the characteristics of microstructures on IC chip surfaces at different granularities. For each subband, we calculated the correlations between the test and reference data and reported the authentication results in terms of EER. The authentication results as a function of the index of subbands are shown in Fig.~\ref{fig:compare_physical_features_cam_sca}. The $v$-shaped black curves in both plots reveal that the sixth subband is the most discriminative physical feature, achieving an EER of $5 \times 10^{-3}$ for Laplacian~(also reported in the third row of TABLE~\ref{tab:comp-main}) and $6 \times 10^{-4}$ for Gaussian. We use two horizontal lines to indicate the performance baselines for comparison. The red dashed horizontal lines refer to the case of using the $y$-component as the feature in Fig.~\ref{fig:camera-scanner_nm}(b). The blue dashed lines refer to the case of using the pixel intensity of raw images as the feature for authentication, achieving EERs of $5 \times 10^{-2}$ under both Laplacian (reported in the first row of TABLE~\ref{tab:comp-main}) and Gaussian assumptions. Both baselines exhibit one order of magnitude worse performance compared to the best subband feature.

\section{Proposed Efficient and Accurate Authentication with Specular Reflection}
\label{sec:fast-acc-auth-spec}

Compared to using diffuse-reflection-based features such as the $x$- or $y$-component of the norm map or a subband of the height map, we demonstrate in this section and Section~\ref{sec:exp} that leveraging a specular-reflection-based feature may be more practically relevant for IC surface authentication---It can reduce authentication error rates by 6.25 to 25 times compared to typical diffuse-reflection-based features while simplifying the imaging setup. This section presents the design of the proposed authentication algorithm and Section~\ref{sec:exp} presents extensive experimental evaluations. 

\subsection{Motivation of Using Specular Points for Authentication}
\label{sec:auth-spec-pts}

Our proposed method of leveraging specular points for authentication is rooted in the everyday observation that a viewer can see glossy spots appear and disappear spontaneously as the viewing angle or the incident light angle on a lit surface changes smoothly. Such surfaces include asphalt roads, flat paper sheets, and IC packages.

One challenge for the camera-based method described in Section~\ref{subsec:auth-feature-diffuse} to be considered viable in real-world supply-chain verification is the requirement for a verification worker to take multiple photos of an IC chip in question. This process may be too slow and stressful, as the worker has to carefully orient the IC and the lamp before taking the desired photos. In fact, while the worker is setting up and before the shutter is pressed, the camera could capture a video containing all the necessary information for verification. Therefore, we favor exploring video-based authentication over photo-based authentication, considering real-world viability.

Our proposed authentication algorithm leverages short video clips of a lit surface with its unique glossy spots and uses their locations for fingerprinting purposes. Our algorithm alleviates the pressures on verification workers, allowing them to take ``lousy" video clips instead of ``perfect" photos, made possible by sampling a small number of frames from the video clip for authentication. From the perspective of detection theory, our test statistic is specially crafted to reduce both false positives [see Eq.~\eqref{eq:Tmax}] and false negatives [see Eq.~\eqref{eq:T_c-score}]. We now formally define the concepts and explain the steps to construct the test statistic supported by pilot data.

\subsection{Definitions}
\subsubsection{Robust specular points}
When specular reflection is observed within a specific working pixel, only a small portion of the area covered by the pixel may have the correctly oriented microscopic surface causing the specular reflection. Since the chip surface is continuous, a small perturbation of the camera or light source location could lead to a shift of the specular reflection to another small portion within the working pixel or the neighboring eight pixels. We name such working pixels that lead to persistent specular reflection \textit{robust specular points}. It is a specular-reflection-based feature that could potentially be exploited for IC chip authentication. 

\subsubsection{Observed specular points}
We formally define a robust specular point in the context of a pair of already-registered test image $\X^\text{t}$ and reference image $\X^\text{r}$.
We define \textit{observed specular points} to be a set of $N$ brightest pixels in the background of a test/reference image. 
A pixel at $(i,j)$ is then called a robust specular point if $X^\text{t} (i,j)$ is an observed specular point, and $X^\text{r} (i,j)$ together with its 8-adjacent neighboring pixels contains at least one observed specular point.

\subsubsection{Robust matching score}
Defining $n(\X^\text{t},\X^\text{r})$ to be the number of robust specular points in $\X^\text{t}$ with reference to $\X^\text{r}$, we can then define a symmetric \textit{robust matching score} as follows:
\begin{equation}
\Srm = \big[ n(\X^\text{t},\X^\text{r} )+n(\X^\text{r},\X^\text{t}) \big] / 2.
\label{eq:robust-matching}
\end{equation}
This robust matching score $\Srm$ is expected to be small when the test and reference images are acquired from different IC chip surfaces that do not share the specular points from the same locations.
In contrast, $\Srm$ is expected to be large when the test and reference images are acquired from the same IC chip surface, and the imaging conditions for the two capturing sessions are similar.
The latter requirement may be too restrictive for a real-world verification scheme.
Next, we design more practical test statistics that relax the requirement for two image-capturing sessions to maintain similar conditions.

\subsection{Design of Robust Test Statistic}
\label{sec:design-test-stat}

\begin{figure}[!t]
\centering
  \vspace{-1mm}
  \hspace{-1mm}
  \subfloat[]{\includegraphics[width=0.48\linewidth]{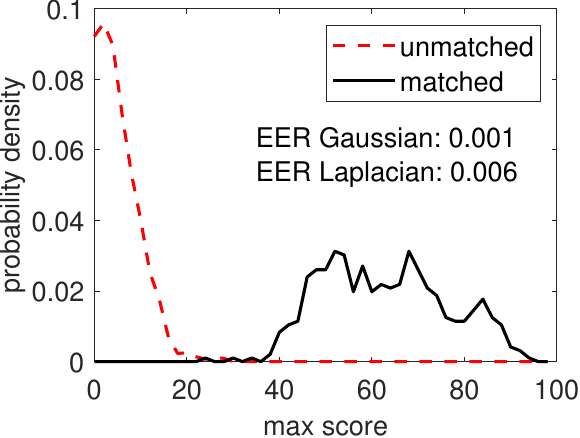}}
  \hspace{2mm}
  \subfloat[]{\includegraphics[width=0.48\linewidth]{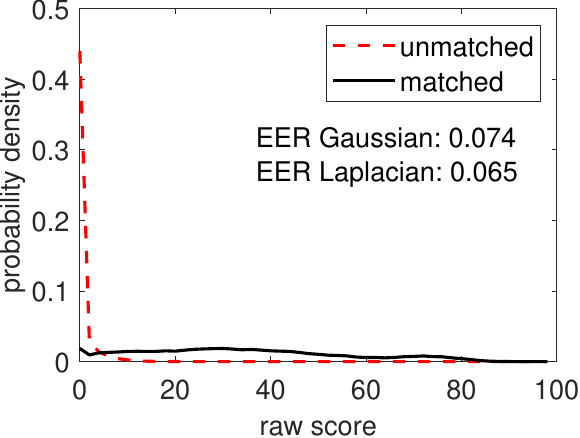}}
  \caption{
  (a) The estimated PDFs under the matched and unmatched cases for the max score $\Tmax$ do not overlap significantly, indicating a reasonable authentication performance due to the use of test--reference frame pairs acquired from the most similar imaging conditions. (b) The PDFs for the raw score $\Srm$ are also plotted for comparison. The wider spread of the PDF of the matched case, caused by mismatched imagining setups of test and reference frames, leads to reduced authentication performance.}
  \label{fig:specular_max_scores}
\end{figure}

For each test video and reference video pair, we randomly sample ten frames from each of the test and reference videos to obtain sample frames that cover a variety of imaging conditions. The robust matching score of every pair of test and reference frames is calculated, resulting in a total of $K = 10 \times 10 = 100$ scores, denoted as $\{S^\text{rm}_i\}_{i=1}^{K}$. 
Because test and reference videos are acquired with a moving light following similar trajectories with procedures detailed in Section~\ref{sec:data-collection}, 
there exist pairs of test and reference frames having large robust matching scores $\Srm$ as they share specular points from the same locations.
Choosing the largest robust matching score as the test statistic is equivalent to the use of the test and reference frames acquired from the most similar imaging conditions, without adding an extra operational burden on the person who conducts the authentication.
Accordingly, we define the \textit{max score} as follows:
\begin{equation}
\Tmax = \max_{i \in \{1, \dots, K\}} S^\text{rm}_i.
\label{eq:Tmax}
\end{equation}
We plot the estimated PDFs under matched and unmatched cases for the max score $\Tmax$ in Fig.~\ref{fig:specular_max_scores}(a).\footnote{We apply the idea of bootstrapping~\cite{stat-bootstrap} to obtain more scores for creating smoother PDFs by repeating the process of randomly sampling frames from the videos $50$ times.}
When using $\Tmax$ as the test statistic with the simple thresholding decision rule, the authentication system can achieve an EER of $6\times10^{-3}$.
Fig.~\ref{fig:specular_max_scores}(b) provides a baseline comparison, when all raw scores $S^\text{rm}_i$ in \eqref{eq:Tmax} before taking the maximum are used, i.e., not requiring test and reference images to have the most similar imaging conditions. As expected, the authentication performance worsens with an EER of $6\times10^{-2}$, mainly due to the increased overlap caused by the wider spread of the PDF of the matched cases.

While the max operation in \eqref{eq:Tmax} leads to an overall improvement in authentication performance, it also has an unwanted byproduct---Fig.~\ref{fig:specular_max_scores}(a) shows a wider spread of the PDF under the unmatched cases as compared to Fig.~\ref{fig:specular_max_scores}(b). 
The design of the test statistic should, therefore, aim to shrink the spread of the PDF under unmatched cases by incorporating more information from $\{S^\text{rm}_i\}_{i=1}^{K}$ in addition to merely taking one of the order statistics.
We approach this design task by setting the max score to zero under those probable unmatched cases in which the max operation wrongly returns a large score due to a few pairs of accidental matches between the test and reference images.
We exploit the fact that under unmatched cases, even with accidental matches, most scores in $\{S^\text{rm}_i\}_{i=1}^{K}$ will be zero as indicated by the red PDF in Fig.~\ref{fig:specular_cus_scores}(b) (previewing here; will be discussed in more detail in Section~\ref{subsec:main-results}).
The \textit{ratio of scores being zero} is defined as follows: 
\begin{equation}
r = \frac{1}{K} \sum_{i=1}^{K} \mathds{1}[S^\text{rm}_i=0],
\label{eq:zero-ratio}
\end{equation}
where $\mathds{1}[\cdot]$ is the indicator function.
On the other hand, not many scores in $\{S^\text{rm}_i\}_{i=1}^{K}$ are zero for matched cases as shown by the black PDF in Fig.~\ref{fig:specular_cus_scores}(b). 
Incorporating the above observations, we define a \textit{robust score} as follows:
\begin{equation}
\Tc = \Tmax \cdot \mathds{1}[r < \tau],
\label{eq:T_c-score}
\end{equation}
\noindent{}where $\tau$ is a tunable threshold and 
$[r<\tau]$ is the random event that less than $(100\cdot \tau)\%$ of test--reference frame pairs achieving absolutely zero in their robust match scores $\Srm$. 
Based on the cross-over between the two curves in Fig.~\ref{fig:specular_cus_scores}(b), we choose to set $\tau = 0.25$ to zero out $\Tmax$ under probable unmatched cases, i.e., $[r \ge \tau]$ and leave $\Tmax$ untouched in probable matched cases, i.e., $[r<\tau]$.
Fig.~\ref{fig:specular_cus_scores}(a) shows the estimated PDFs for the robust scores $\Tc$ under matched and unmatched cases.
There is absolutely no overlap of robust scores between matched and unmatched cases with a significant separation between the PDFs, indicating a further improved authentication performance over the max score $\Tmax$.

\section{Experimental Results}
\label{sec:exp}

In this section, we evaluate the practicality and deployability of the proposed specular-reflection-based authentication method. 
We provide data collection details, assess the accuracy and reliability of the proposed system, and conduct ablation and factor analyses.
The extensive experimental results aim to provide a guide to the optimal mode of operation for the proposed specular-point-based authentication method.

\subsection{Datasets Collection}

\subsubsection{8-Chip Dataset}
\label{sec:data-collection}
We captured a total of 24 videos, three for each of eight distinct memory IC chips, to aid in the algorithm design for the specular-reflection-based authentication method proposed in Section~\ref{sec:fast-acc-auth-spec}.
We used a camera to capture videos for the IC chips, which is a relaxation to the imaging conditions as compared to capturing images in Section~\ref{sec:section-diffuse}.
We fixed a 2018 iPad Pro above each IC chip surface to take videos, which ensures that all chips are captured with consistent orientations, eliminating the need for compensating perspective distortions.
Each chip has a dimension of $0.51~\text{in} \times 0.43~\text{in}$ and an effective resolution of $\sim\!583$~ppi in our 1080p videos.
Before the start of each video, we placed a lamp in front of the IC chip, illuminating the chip at an angle of $\sim \! 45^{\circ}$ from the $z$-axis.
We recorded each video with the light source moving slowly toward the IC chip, where the incident light direction changed gradually to $\sim \! 30^{\circ}$. 
We took three videos for each of the eight IC chips.
Each video lasted about $3$--$4$ seconds, resulting in $\sim \! 100$ frames.
We aligned all video frames using the phase-correlation-based alignment algorithm described in Appendix~\ref{subsec:alignment-phase}.
For matched cases in hypothesis testing, two videos from the same IC chip are used as a test--reference video pair. For unmatched cases, two videos from different IC chips are used as a test--reference pair.

\subsubsection{36-Chip Dataset}

To evaluate the deployability of the specular-reflection-based authentication method, we further collected a larger-scale dataset of 108 videos from 36 distinct chips using a slightly modified capturing setup.
We used the primary camera of an iPhone 13 to capture 4K-resolution videos for the IC chips, each measuring $0.43~\text{in} \times 0.30~\text{in}$, at an effective resolution of $\sim\!1,\!198$~ppi.
Since this is a different type of memory chip than that in the 8-chip dataset, we experimented with different ranges of angles to capture enough specular points.
We chose a broader range from $\sim \! 50^{\circ}$ to $\sim \! 25^{\circ}$. 
For each chip, we captured three $3$--$4$ second videos for a total of 108 videos.
We designed a robust alignment algorithm, detailed in Appendix~\ref{subsec:alignment-sift-gradient}, to accommodate the unique reflection characteristics of the memory chips.

\subsection{Main Results for Specular-Reflection-Based Method}
\label{subsec:main-results}
\begin{figure}[!t]
\centering
  \vspace{-1mm}
  \hspace{-1mm}
  \subfloat[]{\includegraphics[width=0.48\linewidth]{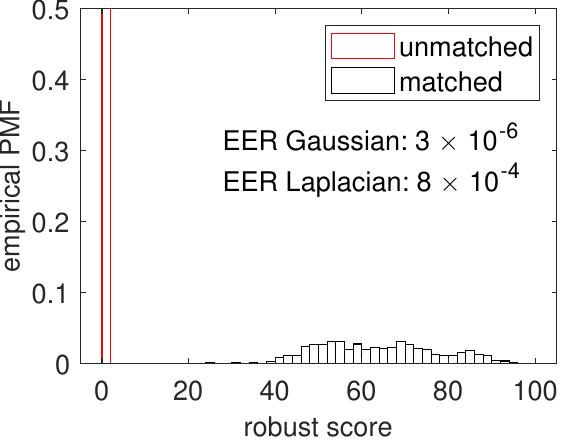}}
  \hspace{2mm}
  \subfloat[]{\includegraphics[width=0.48\linewidth]{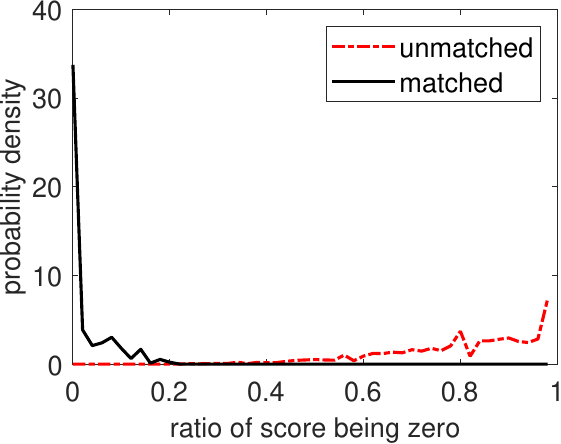}}
  \caption{
  (a)~The estimated PMFs under matched and unmatched cases for the robust scores $\Tc$ are completely separated, indicating further improved authentication performance as compared to that in Fig.~\ref{fig:specular_max_scores}(a). The scores under unmatched cases concentrate at zero, indicating the effectiveness of the multiplicative indicator term in $\Tc$. (b)~The estimated PDFs for the ratio of scores in $\{S^\text{rm}_i\}_{i=1}^{100}$ being zeros \eqref{eq:zero-ratio} under matched and unmatched cases. The fact that most $S^\text{rm}_i$s are zero under unmatched cases and nonzero under matched cases is exploited in the design of the test statistic $\Tc$.}
  \label{fig:specular_cus_scores}
\end{figure}

We first experimentally demonstrate that the proposed robust score $\Tc$, a specular-reflection-based feature, outperforms diffuse-reflection-based features. 
We plotted the PMFs for matched and unmatched cases for the robust score
in Fig.~\ref{fig:specular_cus_scores}(a).
We observe a complete separation of the distributions, leading to an EER of $8 \times 10^{-4}$ assuming that scores follow the Laplace distribution (or $3\times 10^{-6}$ assuming Gaussian).
The superior performance of the robust score is partially attributed to the behavior of the ratio of robust scores being zero, as shown in Fig.~\ref{fig:specular_cus_scores}(b).
For unmatched cases, its PDF is skewed toward one, whereas for matched cases, the PDF is skewed toward zero. 
This occurs because the majority of robust matching scores for unmatched cases are close to zero, whereas those for matched cases are predominantly nonzero.
With this fact incorporated into the design of $\Tc$ through the multiplicative indicator term in \eqref{eq:T_c-score}, the robust score achieves outstanding authentication performance.

TABLE~\ref{tab:comp-main} reveals that the proposed robust score $\Tc$ outperforms all baselines by orders of magnitude in EER. Specifically, the proposed specular-reflection-based method is more reliable than the two diffuse-reflection-based methods tested in Section~\ref{subsec:auth-feature-diffuse}, outperforming the norm map-based method by $25\times$ and the height-map-based method by $6.3\times$. In other words, under the more ubiquitous mobile camera capturing conditions, glossy spots serve as better authentication fingerprints than norm maps and their derived features. 

\subsection{Ablation Study: Components of Test Statistic $\Tc$}

\begin{table}[!t]
    \centering
    \caption{Authentication Performance for Different Test Statistics.}
    \begin{tabular}{llc}
    \toprule 
    & \centering \textbf{Test Statistic} & \textbf{Laplacian EER} $\downarrow$  \\
    \hline
    \textbf{(A)} & Proposed robust score $(\Tc)$ & $0.0008$ \\ 
    \hline
    \textbf{(B)} & (A)$-$Multiplicative indicator $(\Tmax)$ & $0.0056$  \\ 
    \hline
     \textbf{(C)} & (B)$-$Maximum operation $(\Srm)$ &  $0.0649$ \\   
    \bottomrule
    \end{tabular}
    \label{tab:comp-raw-max-cus}
\end{table}
We investigate how authentication performance is affected by removing the multiplicative indicator term and the maximum operation from our proposed robust score $\Tc$~\eqref{eq:T_c-score}. TABLE~\ref{tab:comp-raw-max-cus} shows that the proposed robust score $\Tc$ achieves the best EER of $8 \times 10^{-4}$. This superior performance can be attributed to the complete separation of empirical PMFs under matched and unmatched cases, as illustrated in Fig.~\ref{fig:specular_cus_scores}(a). When the multiplicative indicator term is removed from $\Tc$, the EER worsens by an order of magnitude to $6 \times 10^{-3}$ as the PDF of unmatched cases spreads to overlap with that of matched cases, as shown in Fig.~\ref{fig:specular_max_scores}(a). This experimental result aligns with the rationale for including the multiplicative indicator $\mathds{1}[r < \tau]$ in the test statistics $\Tc$ to zero out any false-positive match between test and reference video clips originating from different IC chips. Further removing the maximum operation from $\Tmax$ results in the worst EER of $6 \times 10^{-2}$, which is one order of magnitude worse than that of $\Tmax$. Fig.~\ref{fig:specular_max_scores}(b) shows an increased spread of the PDF for matched cases. This occurs because the maximum operation ensures that the magnitude of the score is represented by the frame pair sharing the most similar imaging conditions sampled from the test and reference video clips originating from the same IC chip. This helps shape the score distribution for matched cases skewed toward 1.0. Given that $\Tc$ demonstrates the best authentication performance, we use this test statistic in the subsequent ablation studies.

\subsection{Ablation Study: Skipping Specular Points Extraction and Directly Using Raw Frames}
\label{subsec:auth-feature-video-frames}

\begin{figure}[!t]
  \centering
  \vspace{-1mm}
  \hspace{-1mm}
  \subfloat[]{\includegraphics[width=0.48\linewidth]{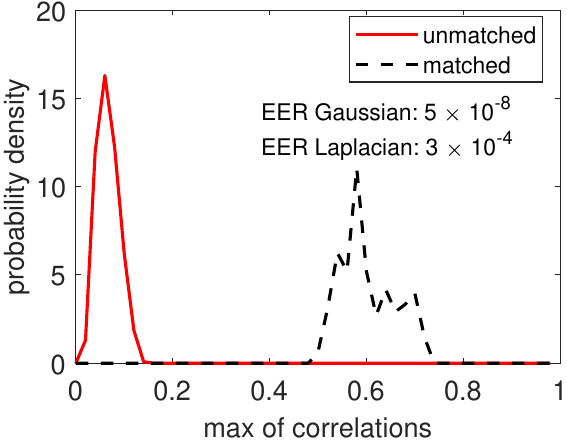}}
  \subfloat[]{\includegraphics[width=0.48\linewidth]{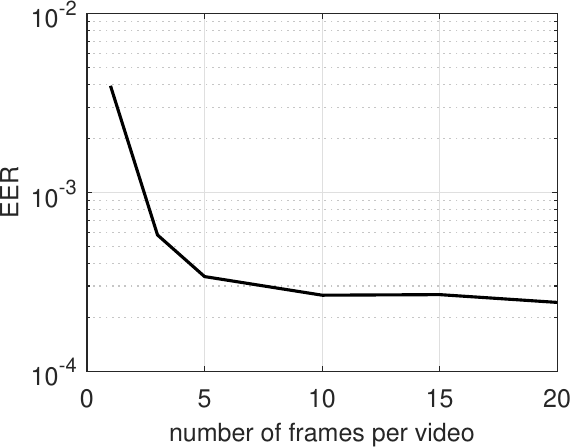}}
  \caption{
  (a)~The estimated PDFs for the max-correlation values under the matched and unmatched cases. It is noted that the empirical PDFs are completely separated with no overlap. 
  (b)~The impact of the number of frames sampled per video on authentication performance, with the EER plotted against the number of frames, assuming a Laplace distribution. Increasing the number of sampled frames results in a better EER, although the improvements eventually plateau.
  }
  \vspace{-4mm}
  \label{fig:correlation_max_scores}
\end{figure}

The authentication method proposed in Section~\ref{sec:fast-acc-auth-spec} identifies locations with consistent specular reflection for IC chip authentication. In this subsection, we examine the effect of bypassing the specular point extraction process by directly authenticating using raw video frames without feature extraction. We demonstrate that using raw video frames as the authentication feature can also achieve excellent performance, but it requires significantly higher storage and communication capacities for verification payloads, thus hindering its practical deployment in large-scale supply chain scenarios.

To conduct the ablation study, we followed a similar procedure outlined in Section~\ref{sec:design-test-stat} to construct a maximum-score-based test statistic, except that we replaced the identification of the locations of observed specular points with the use of the raw video frames. We sampled ten frames from each of the test and reference video clips to form the test and reference sets. Similar to the method of calculating the $\Tmax$ test statistic using the largest robust matching score from $K = 100$ candidates, we calculated $\max(\{c_i\}_{i=1}^{100})$, where $c_i$ is the correlation coefficient between the intensities of the background regions of two randomly sampled test and reference frames. We plotted the PDFs for the matched and unmatched cases for this max-correlation score in Fig.~\ref{fig:correlation_max_scores}(a). It is observed that the empirical distributions are completely separated, leading to a Laplacian EER of $\EditPrasun{3 \times 10^{-4}}$, which is comparable to the performance of the proposed method in Section~\ref{sec:fast-acc-auth-spec}. We also varied the number of selected frames, $\sqrt{K}$, and calculated the Laplacian EER, as shown in Fig.~\ref{fig:correlation_max_scores}(b). While increasing the number of frames improved the EER, the improvements began to plateau around sampling $\sqrt{K} = 10$ frames.

Although this ablation study reveals excellent verification performance using raw video frames as the authentication feature, it has limited practical applicability compared to using specular points as the feature. The specular points approach requires only storing the coordinates of $N$~(usually between 100 and 200, see Section~\ref{subsec:factor-more-frames-n-points}) observed specular points per frame, whereas the raw frame approach requires storing at least one color channel of all pixel values, which could amount to millions of numbers. These payload data also need to be transmitted between a client and server for verification, creating another overhead on data communication. Semiconductor supply chains are vast, processing billions of IC chips monthly \cite{IC_Scale} and involving exchanges between various stakeholders, including manufacturers, assemblers, testers, and consumers. In such complex networks, the payload storage and communication overhead using the raw frame approach could be a significant drawback.

\subsection{Ablation Study: Edge Masking and Detailed Masking}
\label{subsec:ablation-masking}

\begin{figure*}[!t]
\centering
  \includegraphics[]{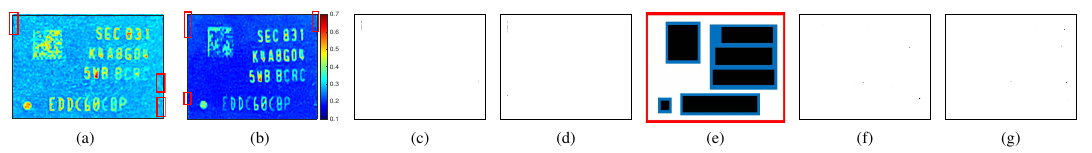}
  \caption{Colormaps of (a)~test frame~49 from chip~\#24 and (b)~reference frame~38 from chip~\#28. 
  The red-box-highlighted glares at the edges are due to the specular reflection at the rounded corners of the chip packaging.
  \textbf{Without edge masking}: The detected glares match at the top-left edges for (c)~chip~\#24 and (d)~chip~\#28, even though the test and reference frames from different chips should not share collocated, observed specular points. The robust matching score is therefore wrongly inflated. 
  (e)~Edge masking (ablation factor~A2, in red) eliminates edge glares, whereas detailed masking (ablation factor~A1, in blue) exposes more background of the chip surface for authentication. Edge masking reduces the EER by 21$\times$ or 1.3 orders of magnitude, while detailed masking achieves a reduction of 1.7$\times$. When both masking tactics are applied together, the EER is reduced by 28$\times$ and 1.5 orders of magnitude.
  \textbf{With edge masking}: Specular points from (f)~chip~\#24 and (g)~chip~\#28 after removing edge glares are now correctly detected. The robust matching score is low for this test--reference frame pair as they belong to different chips.
  [For~(e), all white pixels are never masked. For (c), (d), (f), and (g), the black and white pixels represent the presence and absence of observed specular points, respectively.]}
  \label{fig:mask-with-spec}
\end{figure*}

Our proposed alignment algorithm uses two masking tactics, \textit{edge masking} and \textit{detailed masking}. We investigate the impact of these two tactics on authentication performance and summarize the results in TABLE~\ref{tab:comp-ablation}. As illustrated in Fig.~\ref{fig:mask-with-spec}(e), edge masking (highlighted in red) eliminates pixels along the edges of the IC chip, whereas detailed masking (highlighted in blue) retains the background regions of the chip surface.

\begin{figure}[!t]
\centering
  \vspace{-1mm}
  \hspace{-1mm}
  \subfloat[]{\includegraphics[width=0.47\linewidth]{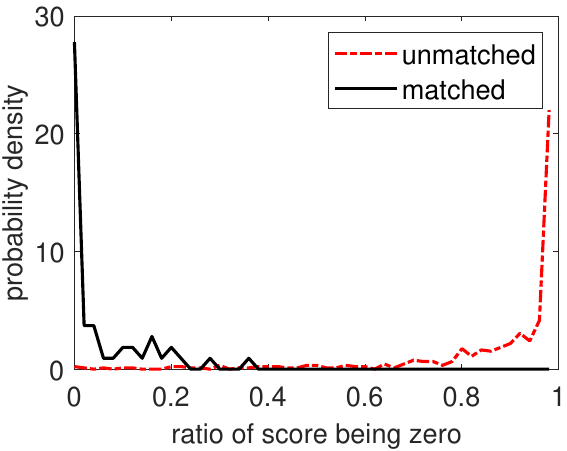}}
  \hspace{2mm}
  \subfloat[]{\includegraphics[width=0.49\linewidth]{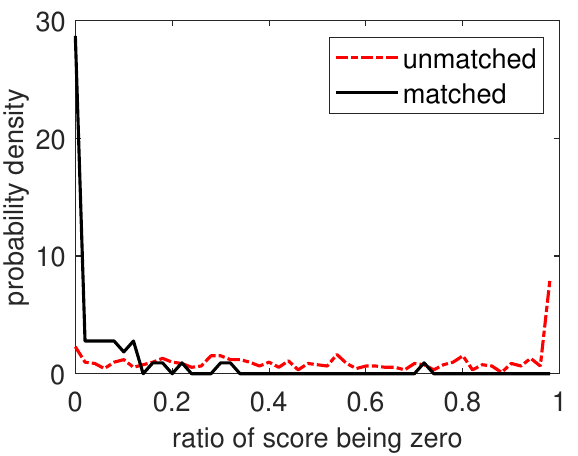}}
  \caption{Estimated PDFs of the ratio of scores being zero with (a)~both detailed and edge masking and (b)~detailed masking only. A significant left uptick in the PDF of unmatched cases in (b) is observed, implying edge masking is effective in reducing the overlap of PDFs.
  }
  \label{fig:ratzero-specularglare}
\end{figure}

TABLE~\ref{tab:comp-ablation} shows that when edge masking (ablation factor A2) is excluded, and only detailed masking is applied, the EER increases significantly from 0.0033 to 0.0528, worsening by about an order of magnitude. Without edge masking, noticeable unintended specular glares along the edges of chips, highlighted in the red boxes in Fig.~\ref{fig:mask-with-spec}(a)--(b), dominate the detected specular points. These falsely matched glares between test and reference frames from unmatched chip units can wrongly inflate the robust matching score, resulting in a left uptick in the PDF of the ratio of scores being zero under unmatched cases, as illustrated in Fig.~\ref{fig:ratzero-specularglare}(b). Ideally, this ratio's distribution should skew toward 1.0 for unmatched cases since test and reference frames from different chips do not share specular points at the same locations. By applying edge masking, the alignment algorithm effectively eliminates the edge glares and reduces the left uptick of the PDF under unmatched cases, as shown in Fig.~\ref{fig:ratzero-specularglare}(a). Thus, edge masking reduces the chance of accidental matches by ensuring that only the central regions, which are less prone to glare-induced artifacts, contribute to the robust matching score.

TABLE~\ref{tab:comp-ablation} reveals that when detailed masking (ablation factor A1) is excluded, and only edge masking is applied, the EER increases slightly from 0.0033 to 0.0044. This increase occurs because excluding detailed masking reduces the available background region [see the blue regions in Fig.~\ref{fig:mask-with-spec}(e)], leading to a reduction in the number of matched specular points in test--reference frame pairs. Keeping the background regions between text lines provides additional specular points for these frame pairs, ensuring that the robust matching scores are nonzero. This, in turn, helps prevent zero-valued robust matching scores for matched cases.

When both masking (ablation factors A1 \& A2) tactics are excluded, we obtain the worst EER of 0.0920. Compared to the EER of 0.0528 when only edge masking is removed, the significantly worsened EER reveals a strong interaction between the two masking tactics.

\begin{table}[!t]
    \centering
    \caption{Authentication Performance Affected by Masking Tactics}
    \begin{tabular}{lc}
    \toprule 
    \centering \textbf{Ablation Configuration} & \textbf{EER for} $\Tc$  \\
    \hline
    Proposed Method  & $0.0033$\\
    \hline 
    \hline
    Excl. (A1)~Detailed Masking &  $0.0044$ \\ \hline  
    Excl. (A2)~Edge Masking & $0.0528$  \\ \hline  
    Excluding (A1) \& (A2) & $0.0920$ \\
    \bottomrule
    \end{tabular}
    \label{tab:comp-ablation}
\end{table}

\begin{figure}[!t]
\centering
  \vspace{-1mm}
  \hspace{-1mm}
  \subfloat[]{\includegraphics[width=0.48\linewidth]{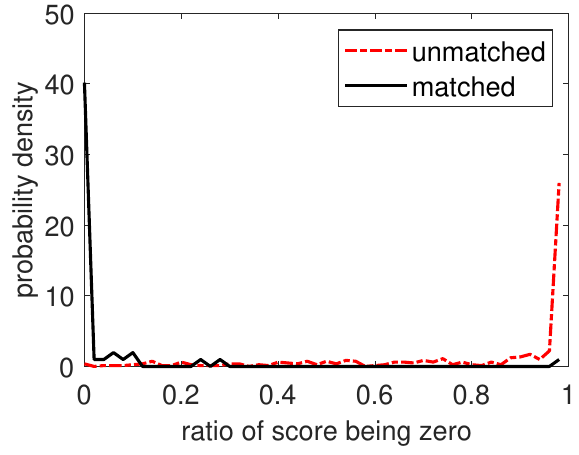}}
  \hspace{2mm}
  \subfloat[]{\includegraphics[width=0.48\linewidth]{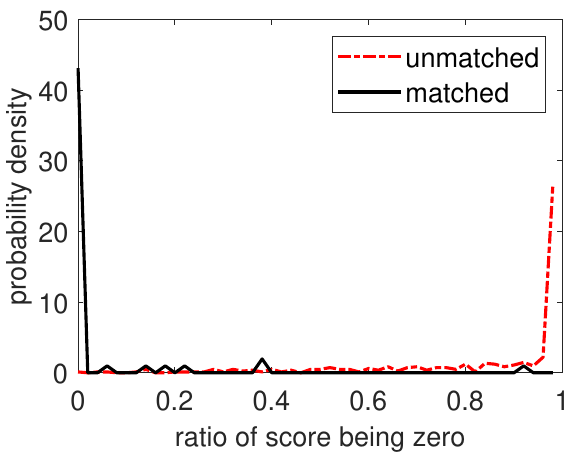}}
  \caption{Estimated PDFs for ratio of scores being zero for random seeds (a)~``123" and (b) ``456". The random seed alters the details of PDFs but not their shapes.
  This stability in distributions of the ratio of scores being zero contributes to the robustness of the proposed test statistic~$\Tc$ against random frame sampling, ensuring that authentication performance varies within the same order of magnitude.}
  \label{fig:rnd-seed-ratzero}
\end{figure}

\subsection{Factor Analysis: Random Frame Sampling}
\label{subsec:seed-selection}

We investigate the impact of using random seeds during the frame sampling stage on authentication performance, demonstrating that the variation remains within the same order of magnitude. We utilized chips \#1 to \#18 to analyze the effect of random frame sampling and reported the EER for different random seeds in TABLE~\ref{tab: rnd-seed-auth}.
\begin{table}[!t]
    \centering
    \caption{Authentication Performance Under Various Random Seeds}
    \begin{tabular}{cc}
    \toprule 
    \textbf{Random Seed} & \textbf{EER for} $\Tc$ \\
    \hline
    123 & $0.0037$\\ \hline
    456 & $0.0012$\\ \hline 
    789 & $0.0037$\\ \hline 
    101112 & $0.0012$\\ \hline
    131415 & $0.0061$\\ \hline\hline
    \textit{Average} & $0.0032$ \\ \hline
    \textit{Standard deviation} & $0.0021$ \\ \bottomrule
    \end{tabular}
    \label{tab: rnd-seed-auth}
    \vspace{-4mm}
\end{table}
We observed that the EER varies from 0.0012 to 0.0061, with a standard deviation of 0.0021, indicating stability within the same order of magnitude. This result can be attributed to two factors. First, the magnitude component of the robust score is relatively stable. Each random seed yields a unique set of robust matching scores $\{S^\text{rm}_i\}_{i=1}^K$ calculated from the $K$ randomly sampled test--reference frame pairs. Although individual scores in different sets exhibit large variance, it can be shown that the largest order statistic $\Tmax$ has the same mean across different random draws and a substantially reduced variance. Since $\Tmax$ serves as the magnitude component of the robust score $\Tc$, we conclude that random frame sampling introduces only minor perturbations. Second, the indicator component of the robust score is also relatively stable. Fig.~\ref{fig:rnd-seed-ratzero} reveals that the PDFs of the ratio of scores being zero under matched and unmatched cases remain approximately consistent\footnote{The minor inconsistency across the PDFs~(black curves) around 0.4 for matched cases represents an outlying matching case. In this instance, 40\% of the frame pairs did not have a single matched robust specular point.} across two different random seeds. 

Given that the indicator component, with a fixed $\tau = 0.25$, is a conditional Bernoulli random variable based on one of the PDFs of the ratio of scores being zero, the stability of PDFs against random seeds ensures the stability of the indicator component.

\subsection{Factor Analysis: Will More Frames or Observed Specular Points Help?}
\label{subsec:factor-more-frames-n-points}

\begin{figure}[!t]
\centering
  \vspace{-1mm}
  \hspace{-1mm}
  \subfloat[]{\includegraphics[width=0.48\linewidth]{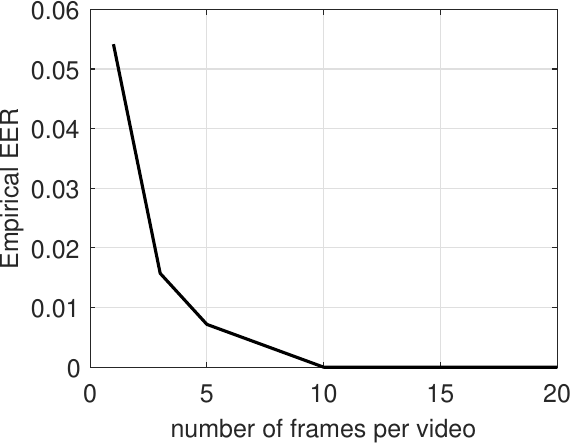}}
  \hspace{2mm}
  \subfloat[]{\includegraphics[width=0.48\linewidth]{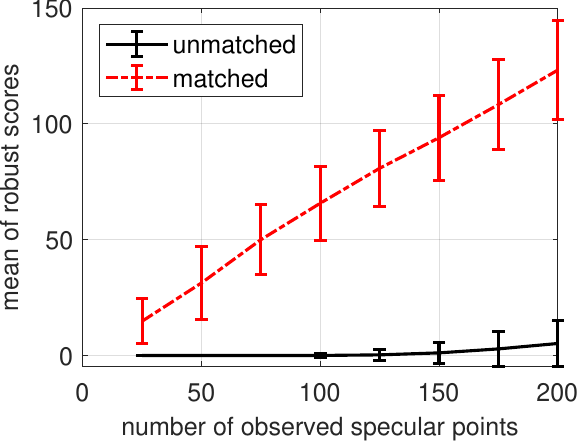}}
  \caption{(a) The impact of the number of frames sampled from each video on the authentication performance in terms of EER. A larger number of frames used will lead to better performance. (b) The impact of the number of frames sampled from each video on the authentication performance. The error bars indicate one sample standard deviation above and below the averaged robust scores. As the number of observed specular points increases, scores for the matched case will increase fast.}
  \label{fig:specular_cus_scores_frame_num}
\end{figure}%

We examine the impact of varying the number of randomly sampled frames. Fig.~\ref{fig:specular_cus_scores_frame_num}(a) illustrates the EER as a monotonically decreasing function of the number of frames sampled per video ranging from $\sqrt{K} = 1$ to $20$ while fixing the number of observed specular points $N = 100$. Meanwhile, the storage and communication costs increase linearly with the number of frames. As the improvement in EER saturates beyond $\sqrt{K} = 10$, selecting 10 to 15 frames per video may be reasonable.

We also examine the impact of varying the number of observed specular points. Fig.~\ref{fig:specular_cus_scores_frame_num}(b) illustrates the evolution of distributions for robust scores (in terms of mean plus and minus standard deviation) for matched and unmatched cases while keeping the number of sampled frames fixed at $\sqrt{K} = 10$. An interesting observation is that even though the mean of the unmatched distribution hardly increases, its standard deviation increases rapidly beyond $N = 125$. This rapid increase could be attributed to false positive matches of frames sourced from two different IC chips, due to wrongly classifying extremely bright diffuse points as specular points. Thus, using $N = 100$ observed specular points may be a reasonable hyperparameter choice, given the clear gap between the two distributions and the small standard deviation of the unmatched distribution.

\section{Conclusion}

In this paper, we have conducted an initial investigation into enabling PUF-based authentication for IC chip surfaces. Our experiments confirm that consumer-grade scanners and cameras can effectively capture meaningful diffuse-reflection-based physical features of IC chip surfaces for authentication. By leveraging stable specular points, we have proposed an effective and lightweight IC chip verification scheme better suited for practical deployment. We have also conducted extensive factor, sensitivity, and ablation studies to understand the detailed characteristics of the proposed lightweight verification scheme, aiming to guide real-world deployment. This research paves the way for applying optical PUF techniques, synergizing image and video processing with circuits and systems research, to verify individual IC chip units, complementing existing electronic PUF techniques in combating counterfeit activities in supply chains. 

~\\
\noindent{}\textbf{Acknowledgment.} The authors thank Nayeeb Rashid for his effort in collecting the initial set of videos, Jiawei Gao for providing preliminary theoretical analysis results, and Kai Yue for providing constructive review comments.

\begin{appendices}
    \section{Alignment Algorithms}

\subsection{Phase Correlation Based Method}
\label{subsec:alignment-phase}
We propose to align the captured test images of IC chips using phase correlation \cite{reddy1996fft}.
First, we obtain a template image for each IC chip. We scan the surface of the IC chip using a flatbed scanner, which will generate a better-quality image with less noise than using a mobile phone. We then crop the area of interest of the captured image as the template image.
Next, we register the captured test image.
Given the test and template images, we first use phase correlation and Fourier properties \cite{reddy1996fft} to estimate the translational, rotational, and scale movement to determine a geometric transform $T$. Finally, we use $T$ to warp the test image with the template image as a reference to obtain the final registered image. 

\subsection{SIFT and Optimization Based Method}
\label{subsec:alignment-sift-gradient}

\begin{figure}[!t]
\centering
  \vspace{-3.5mm}
  \subfloat[]{\includegraphics[width=0.48\linewidth]{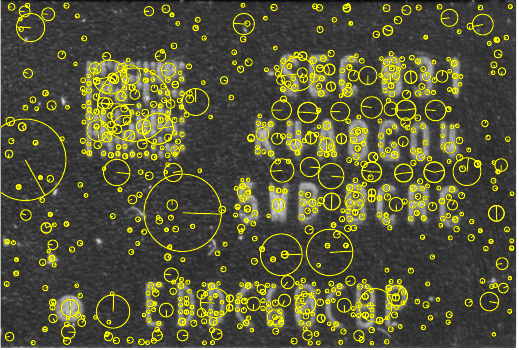}}
  \hfil
  \subfloat[]{\includegraphics[width=0.48\linewidth]{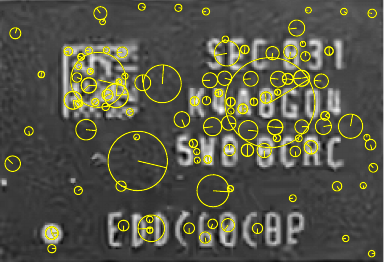}}
  \caption{SIFT keypoints of chip~\#2 extracted from (a) the scanned template image and (b) a frame of the test camera video. }
  \label{fig:ls_sift}
\end{figure}

\begin{figure}[!t]
\centering
  \vspace{-3.5mm}
  \subfloat[]{\includegraphics[width=0.3\linewidth]{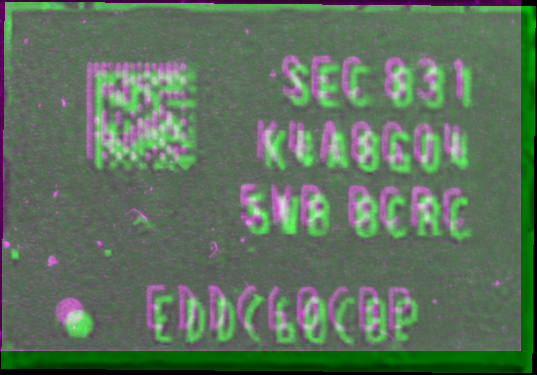}}
  \hfil
  \subfloat[]{\includegraphics[width=0.31\linewidth]{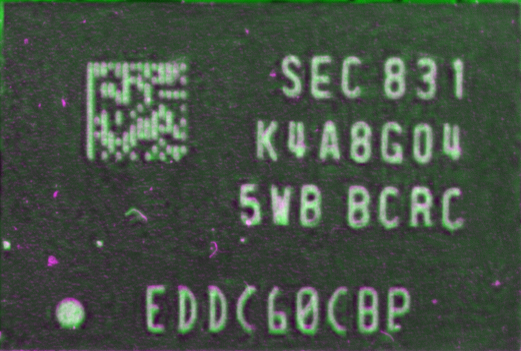}}
  \hfil
  \subfloat[]{\includegraphics[width=0.31\linewidth]{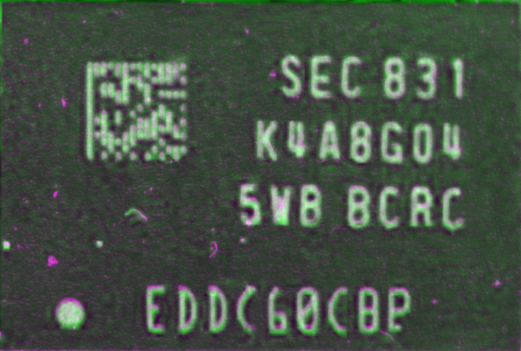}}
  \caption{Overlay of the aligned frame (in green) and the template image (in pink) for chip~\#2 after (a) the SIFT-based alignment, (b) the initial \texttt{imregtform} alignment, and (c) the final \texttt{imregtform} alignment. Note the misalignment between the test frame and the template image when using SIFT-based alignment, and the subsequent improvement after refinement with \texttt{imregtform}.}
  \label{fig:ls_alignment_overview}
\end{figure}
We propose a robust two-stage approach for aligning captured test video frames of IC chips in large-scale IC authentication. This approach combines SIFT~\cite{lowe1999object} with optimization-based techniques~\cite{mattes2001nonrigid, styner2000parametric} to enhance alignment accuracy and robustness.  
First, we obtain a template image for each IC chip using a flatbed scanner, which produces a higher-quality image with less noise compared to using a mobile phone.
We then crop the area of interest from the captured image to serve as the template. 
Next, we align the frames of the captured test video to the template image.
Similar to the template creation process, we crop the area of interest in the video to extract the region corresponding to the specific chip. 
We use the first frame of the cropped video as the test frame to estimate the geometric transformation, which is 
subsequently applied to the remaining frames of the test video.

\begin{figure}[!t]
\centering
  \vspace{-3.5mm}
  \subfloat[]{\includegraphics[height=0.12\textheight,width=0.47\linewidth,trim={0 0 2cm 0},clip]{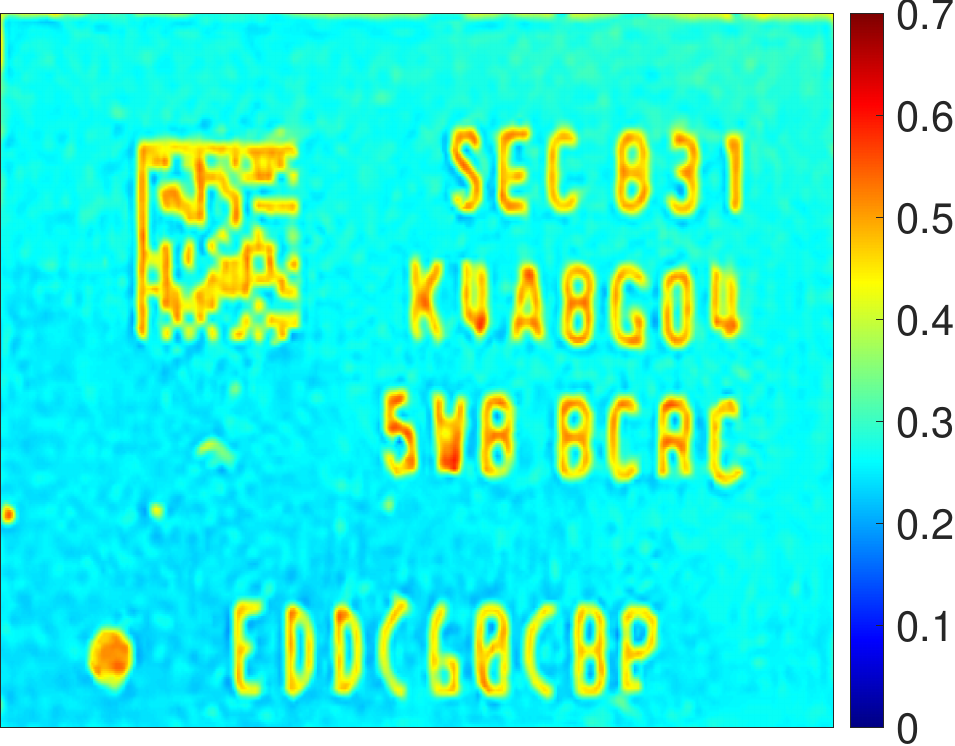}}
  \hfil
  \subfloat[]{\includegraphics[height=0.12\textheight,width=0.51\linewidth]{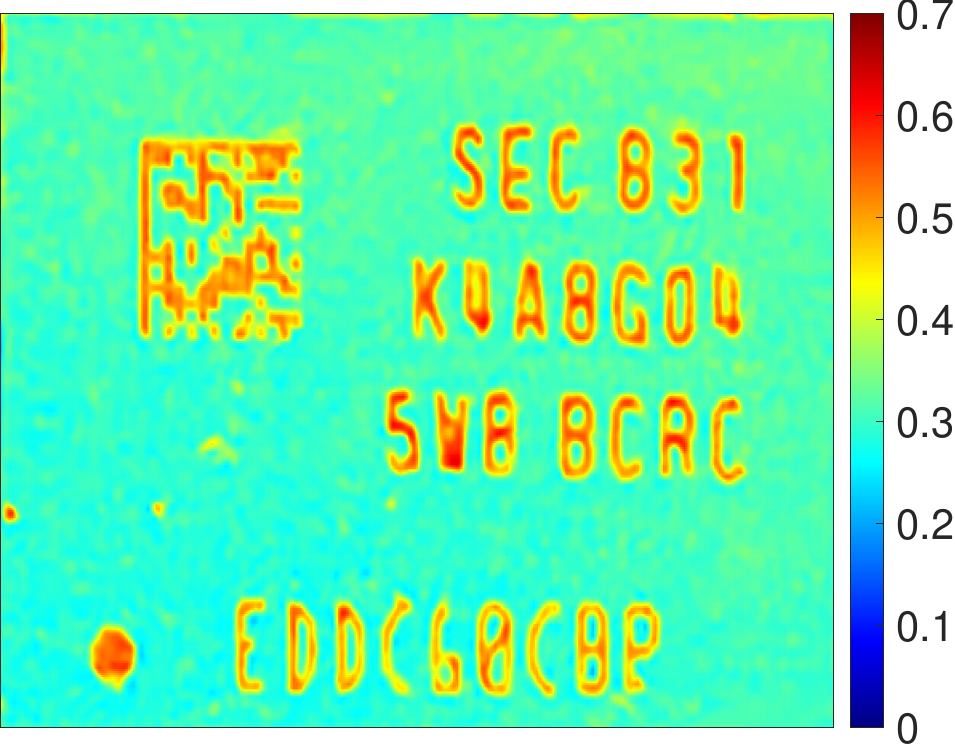}}
  \vspace{-2mm}
  \subfloat[]{\includegraphics[height=0.12\textheight,width=0.47\linewidth,trim={0 0 2cm 0},clip]{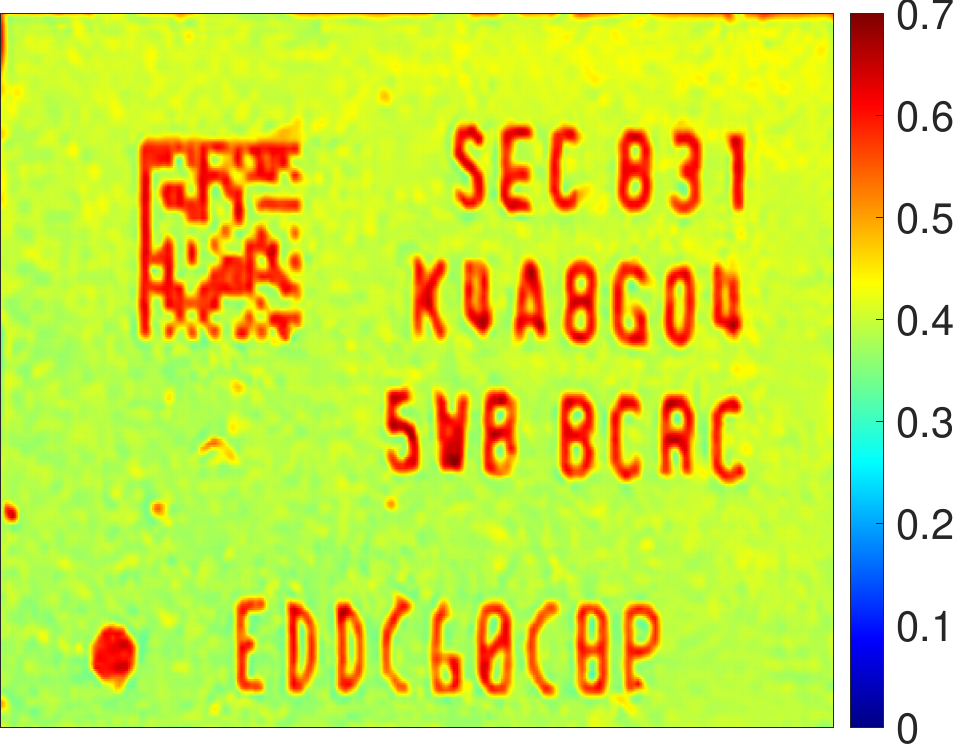}}
  \hfil
  \subfloat[]{\includegraphics[height=0.12\textheight,width=0.51\linewidth]{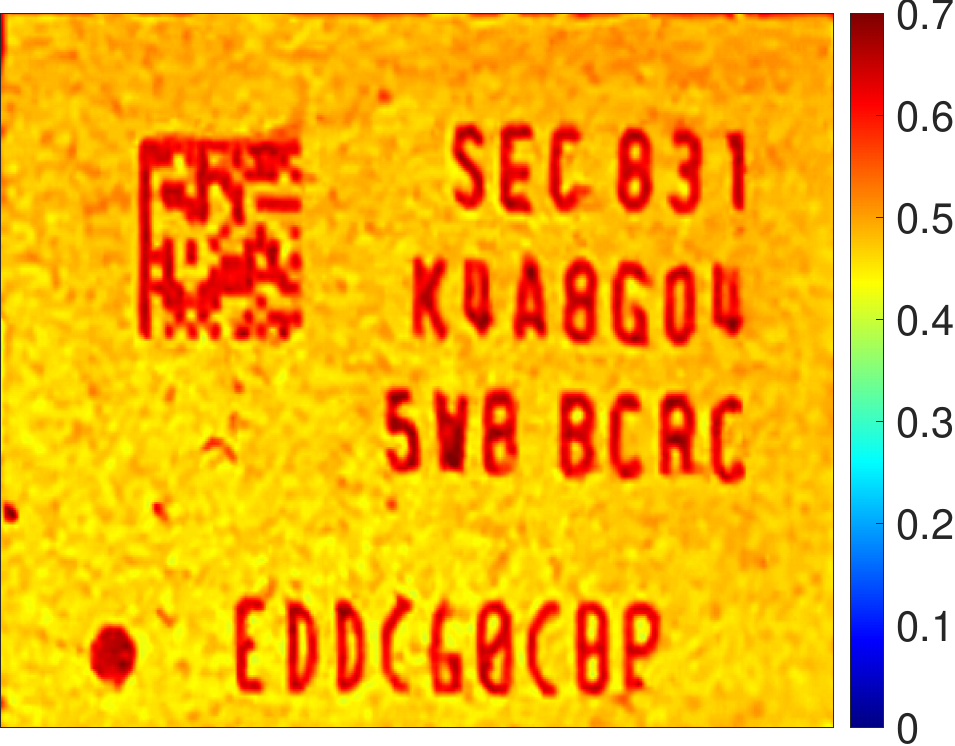}}
  \caption{Aligned frames of test video of chip~\#2 at (a) 1~s, (b) 2~s, (c) 3~s, and (d) 4~s marks. The specular points become 
  increasingly prominent over time as the light source moves closer to the chip.}
  \label{fig:ls_registered_video}
\end{figure}

Given the test frame and template image, we first identify and match their respective SIFT features \cite{lowe1999object} as shown in Fig.~\ref{fig:ls_sift}. 
We then use the matched feature pairs to estimate the translational, rotational, scale, and shear movements to determine the geometric transform $T$. 
After this alignment stage, the test frame and template image are better aligned based on their features, though slight scale, rotation, or translation imperfections may still remain, as shown in Fig.~\ref{fig:ls_alignment_overview}. 
To ensure precise alignment, we further refine the process using intensity-based alignment through \textsc{Matlab}'s built-in \texttt{imregtform} function.
We utilize the function's ``multimodal" setting as the template image is acquired 
via a scanner while the test video is recorded 
using a mobile camera.
In this setting, the \texttt{imregtform} function optimizes Mattes mutual information criterion \cite{mattes2001nonrigid} across multiple image pyramid levels using an evolutionary optimization technique \cite{styner2000parametric}.
Using the high-quality initialization $T$, we learn a 
refined geometric transform $\tilde{T}$ to estimate the precise translational, rotational, scale, and shear movements by running the optimization twice.
Running the optimization twice with progressively finer search radii helps in convergence and 
results in a more precise alignment.
Finally, $\tilde{T}$ is used to warp frames of the test video with the template image as a reference to obtain the frame-wise aligned video, as shown in Fig.~\ref{fig:ls_registered_video}.

    \section{Norm Map Estimation Algorithm for Surfaces with Both Diffusion and Specular Components}
\label{sec:norm-map-est-alg}

In the main paper, we followed the norm map estimation procedure outlined in Liu and Wong~\cite{liu2021microstructure} to calculate norm maps for IC chip surfaces that contain both diffuse and specular reflection components. Specifically, we scanned each IC chip surface in two pairs of opposite directions and took the difference between the scanned images of each pair to obtain the scaled $x$- or $y$-component of the norm map. Although this estimation algorithm is the same as the one for fully diffuse surfaces proposed by Clarkson et al.~\cite{clarkson-09}, the derivation in Liu and Wong~\cite{liu2021microstructure} is more general. To make the paper more self-contained, we summarize the key steps of their derivation below.

Under the generalized reflection model that encompasses both diffuse and specular reflection~\cite{book}, the perceived intensity $l_r$ at location $\p$ may be modeled as follows:

\begin{equation}
l_r = \frac{l}{\|\o-\p\|^2} \Big[ w_d  \cdot (\n^T\v_i)^+ + w_s  \cdot (\v_c^T \v_r)^{k_e} \Big],
\label{sp}
\end{equation}
\noindent{}where $\n=(n_x, n_y, n_z)^T$ is the microscopic normal direction of the paper surface at location $\p$, $\o=(o_x, o_y, o_z)^T$ is the location of the light source,  $\v_i = (\o-\p) / \| \o - \p \|$ is the incident light direction, $l$ is the strength of the light,
$1/\|\o-\p\|^2$ is a light-strength discounting factor as the received energy per unit area from a point light source, which is inversely proportional to the squared distance.
$x^+ = \max(0,x)$, and
$k_e>0$ controls the gloss level of the surface.
$w_d$ and $w_s$ are the weights for diffuse and specular components, 
$\v_c$ is the camera's direction, and $\v_r$ is the specular reflection direction.

The pixel value at location $\p$ in the image of a scanner can be expressed as the integral over all light diffusely and specularly reflected off that surface point, originating from points $\o$ along the linear light path in the $x$-direction \cite{liu2021microstructure}:
\begin{equation}
\begin{split}
I_{0^\circ} & = \int_{x_1}^{x_2} l_r do_x  = l \int_{x_1}^{x_2} \big( w_d\n^T\v_i + w_s\v_c^T\v_r \big)\frac{1}{\|\o\|^2} \ do_x \\
\end{split}
\end{equation}

Under the generalized reflection model, scanning the chip surface in two opposite directions to acquire images $I_{0^\circ}$ and $I_{{180}^\circ}$ and computing $I_{0^\circ} - I_{{180}^\circ}$ is still proportional to $n_y$ at location $\p$, as theoretically shown by Liu and Wong \cite{liu2021microstructure}:
    
\begin{equation}
\begin{split}
&I_{0^\circ} - I_{{180}^\circ} = sn_y  \\
&+ 2l \int_{x_1}^{x_2} \bigg[ w_s\v_c^T(\n\n^T-\n'\n'^T)\v_i \bigg] \frac{1}{\|\o\|^2} \ do_x. \\
& \approx \left[ s+  2(v_{cz} +v_{cy} o_z / o_y) s' \right]n_y
\end{split}
\label{eq:scanner_diff}
\end{equation}
\noindent{}where $s=2l w_d o_y (x_2-x_1)$ and $s' = 2l \cdot w_s o_y\int_{x_1}^{x_2}{\|\o\|^{-3}}do_x$. 
\end{appendices}

\bibliographystyle{IEEEtran}
\bibliography{refs_IC_PUF}

\end{document}